\documentclass[graybox, secnum]{svmult}

\usepackage{amsmath,amssymb}
\usepackage{mathptmx}       
\usepackage{helvet}         
\usepackage{courier}        
\usepackage{type1cm}        
\usepackage{makeidx}         
\usepackage{graphicx}        
\usepackage{multicol}        
\usepackage[bottom]{footmisc}
\usepackage{hyperref}        
\usepackage{soul}            
\hypersetup{colorlinks=true,urlcolor=blue}
\usepackage[square,numbers]{natbib}
\makeindex             

\newcommand{\be}{\begin{equation}}\newcommand{\ee}{\end{equation}}
\newcommand{\bea}{\begin{eqnarray}}\newcommand{\eea}{\end{eqnarray}}
\newcommand{\nn}{\nonumber}\newcommand{\p}[1]{(\ref{#1})}
\newcommand{\lb}[1]{\label{#1}}

\newcommand{\HA}{{\Bbb H}\hskip-4.1pt {\Bbb A}}

\def\theequation{\arabic{section}.\arabic{equation}}
\begin{document}
\title*{
${\cal N}=2\,$ Supergravities in Harmonic Superspace}
\author{Evgeny Ivanov}
\institute{Bogoliubov Laboratory of Joint Institute for Nuclear Research, 141980 Dubna, Moscow region, Russia; Moscow Institute of Physics and Technology, 141700 Dolgoprudny, Moscow region, Russia, \email{eivanov@theor.jinr.ru}}
%
%
\renewcommand{\theequation}{\thesection.\arabic{equation}}
\maketitle
\abstract{Basics of ${\cal N}=2, 4D$ conformal and Einstein supergravities in the harmonic superspace approach are outlined. The crucial merit
of this formulation consists in that the relevant off-shell supermultiplets, in particular ${\cal N}=2, 4D$ superconformal Weyl multiplet, are accommodated by the harmonic-analytic unconstrained prepotentials
with a clear geometric meaning,
like in the analogous formulation of ${\cal N}=2, 4D$ supersymmetric gauge theory. The fundamental gauge group of conformal supergravity  is constituted by the analyticity-preserving
diffeomorphisms of harmonic superspace. The superfield actions of various off-shell versions of ${\cal N}=2$ Einstein supergravity are obtained  as the actions of the appropriate harmonic analytic compensators
in the background of conformal ${\cal N}=2$ supergravity. The version admitting  the most general couplings to quaternion-K\"ahler matter  is the ``principal'' version with the unconstrained harmonic analytic hypermultiplet
superfield as a compensator. It involves an infinite number of auxiliary fields.}

\section*{Keywords}
Supersymmetry, Harmonic superspace, Supergravity.
\begin{flushright}
{\it In Memory of V.I. Ogievetsky and A.S. Galperin}
\end{flushright}

\section{Introduction}
The natural geometric framework for  supersymmetric field theories \cite{susy1}-\cite{susy3} is provided by {\it superspace},
an extension of Minkowski space (or, of any other bosonic space)
by anticommuting fermionic (Grassmann) coordinates \cite{susy2,SS1,SS2}\footnote{These variables were treated in \cite{susy2} as fermionic fields given on Minkowski space (Goldstone fermions), while
in \cite{SS1} as independent new anticommuting coordinates.}. The fields defined on superspace
are called {\it superfields} \cite{SS1,SS2}. They naturally describe the supermultiplets
of given supersymmetry. The fields forming these supermultiplets come out as coefficients
in the expansion of the superfields over Grassmann coordinates.
The basic merit of the off-shell superfield approach is the opportunity to formulate the
supersymmetric theories in a systematic and consistent way, making manifest
their non-trivial intrinsic geometries (e.g. the complex geometry of ${\cal N}=1$
supergravity \cite{OS}) and their remarkable quantum properties (e.g.,  the
ultraviolet finiteness of ${\cal N}=4$ super Yang-Mills theory \cite{N4YM}).

The superfield approach to ${\cal N}=1, 4D$ Poincar\'e
supersymmetry was proposed some  fifty years ago in the pioneering papers
\cite{SS1,SS2,Ferr}. It took much
longer to work out a suitable superfield formalism for extended supersymmetries
(i.e. those containing more than one spinor generator).
Even in the simplest case of ${\cal N}=2$ supersymmetry, up to 1984 it was unknown
how to formulate the relevant theories off shell, in a manifestly
supersymmetric form and in terms of unconstrained superfields.
The breakthrough came about with the invention of a new type of
superspace, the harmonic superspace (HSS)  \cite{aG5}-\cite{aG8}, as
a development of the important concept of Grassmann analyticity \cite{GA}.
It  allowed to construct off-shell unconstrained formulations for all the ${\cal N}=2$
supersymmetric theories (${\cal N}=2$ matter, Yang-Mills and
supergravity) and for ${\cal N}=3$ supersymmetric Yang-Mills theory.

Harmonic ${\cal N}=2$ superspace (HSS) is an extension of the standard ${\cal N}=2$ superspace
by the two-dimensional sphere $S^2\sim SU(2)/U(1)$. In such an
extended superspace a new kind of
Grassmann analytic subspace exists, the harmonic analytic one, parametrized by half of the original spinor
coordinates \cite{aG5,aG6,Book}. The Grassmann harmonic analyticity is
the key to finding the adequate off-shell
unconstrained formulations mentioned above, just like ${\cal N}=1$ chirality,
the simplest case Grassmann analyticity, forms the basis of the unconstrained superfield
formulations of ${\cal N}=1$ supersymmetric theories.

The ultimate goal of the present review paper is to give an account of the basic elements of the HSS formulation of
${\cal N}=2, 4D$ supergravity (SG), though we  will also touch some other aspects of the HSS approach.

Sect. 2 contains the introductory information  about supersymmetry and superspaces.
In Sect. 3 the main motivations for HSS are explained
and the basic concepts and technical tools of this approach are described. Sect. 4 collects the
HSS formulations of  ${\cal N}=2$ matter and super Yang-Mills theory as a necessary preparatory step to various versions
of ${\cal N}=2$ supergravity. In Sect. 5 we ``from scratch'' describe the HSS formulation of the
simplest (and historically first) Einstein ${\cal N}=2$ supergravity, basically following ref. \cite{GS1}. We also explicitly
present its linearized superfield action, based on \cite{Zup} and \cite{BIZ1}. In Sect. 6 we describe the HSS formulation
of conformal ${\cal N}=2$ SG \cite{aG16} and then explain, basically on the example of Sect. 5,  how to pass to various versions
of Einstein ${\cal N}=2$ SG by adding the appropriate HSS compensating superfields in the background of conformal SG. We also describe ${\cal N}=2$ supergravity--matter couplings
and point out that the most general coupling can be achieved only in the framework of the so called ``principal'' version of  ${\cal N}=2$ SG using the off-shell
hypermultiplet as a compensator and so involving an infinite number of auxiliary fields. The concluding Sect. 7 gives a brief summary of applications of the HSS approach to
supergravities and  some related theories, including quite recent applications to the off-shell description of ${\cal N}=2$ supersymmetric higher spins \cite{BIZ1}. Remarkably,
the HSS formulation of the latter theory for arbitrary superspin ${\bf s} > 2$  reveals a great resemblance with that  of ${\cal N}=2$ SG presented in Sect. 5, being its
rather straightforward generalization. There is also given an (incomplete) list of problems still waiting their resolution within the HSS approach and its  proper modifications.

In this paper, I mainly focus on the HSS approach to ${\cal N}=2$ theories and apologize for an inevitable incompleteness of the reference list. The more comprehensive list
of references to the component and superfield formulations of ${\cal N}=2$ supergravity can be found, e.g., in \cite{FVP,KRTM}.

\setcounter{equation}{0}
\section{Superspace: basic concepts}

\subsection{${\cal N}$-extended Poincar\'e supersymmetry and superspaces}
The ${\cal N}=1$ Poincar\'e supersymmetry, along with the standard
Poincar\'e group generators $P_m, L_{[m,n]}$ ($m, n = 0,1,2,3$; $P_m$ are the
4-translation generators and $L_{[m,n]}$ the Lorentz group ones), involves the
fermionic Weyl generators $Q_\alpha, \bar Q_{\dot\alpha}\; (\alpha, \dot\alpha
= 1,2)$ which transform as $(1/2,0)$ and $(0,1/2)$ of the Lorentz group and
obey the following anticommutation relations: \be \{Q_\alpha,
\bar Q_{\dot\alpha}\} = 2(\sigma^m)_{\alpha\dot\alpha} P_m\,, \quad \{Q_\alpha,
Q_{\beta}\} = \{\bar Q_{\dot\alpha}, \bar Q_{\dot\beta}\} =0\,, \quad
(\sigma^m)_{\alpha\dot\alpha} = (1, {\vec{\sigma}})_{\alpha\dot\alpha}\,.\lb{N1}
\ee ${\cal N}>1$ extended supersymmetry involves ${N}$ copies of the fermionic
generators, each satisfying relations \p{N1} \be \{Q_{\alpha}^{i}, \bar
Q_{\dot\alpha\,k}\} = 2 \delta^i_k (\sigma^m)_{\alpha\dot\alpha} P_m\,, \quad
\{Q_{\alpha}^{i}, Q_{\beta}^{k}\} = \{\bar Q_{\dot\alpha\,i}, \bar
Q_{\dot\beta\,k}\} =0\,.\lb{N2} \ee
Here $i = 1, \ldots N$ is the index of the
fundamental representation of the internal automorphism symmetry (or
R-symmetry) group $U(N)\,$\footnote{Some important theories, e.g., ${\cal N}=4$ super Yang-Mills
theory, in fact respect only $SU(N)$ R-symmetry.}.

The natural way to realize ${\cal N}$-extended
Poincar\'e supersymmetry is to use the standard superspace \cite{SS1,SS2}
\begin{equation}
{\Bbb R}^{4\vert 4N}=(x^a, \theta^{\alpha}_i\;,
\bar{\theta}^{\dot\alpha i})\;, \hspace*{1cm}  i=1,2,\ldots,N
\label{1.3}
\end{equation}
involving the spinor {\it anticommuting} coordinates
$\theta^{\alpha}_i\;, \bar{\theta}^{\dot\alpha i}$, in addition to the commuting
$x^{a}$. Their transformation rules under the Poincar\'e group are
evident, while the transformations under supersymmetry
(supertranslations with anticommuting parameters
$\epsilon^{\alpha}_i\;,\bar\epsilon^{\dot\alpha i}$) are given by
\begin{equation}
\delta x^a=i(\epsilon^i \sigma^a \bar\theta_i - \theta^i \sigma^a
\bar{\epsilon}_{i})\;, \qquad  \delta\theta^{\alpha}_i =
\epsilon^{\alpha}_i\;,\hspace*{1cm} \delta \bar\theta^{\dot\alpha
i} = \bar{\epsilon}^{\dot\alpha i}\;. \label{1.4}
\end{equation}
Superfields $\Phi(x,\theta,\bar\theta)$ are defined as functions
on this superspace and their transformation law is completely
determined by the superalgebra \p{N2}. For example, for a scalar superfield
\begin{equation}\label{1.5}
\Phi'(x',\theta',\bar\theta')=\Phi(x,\theta,\bar\theta)\;.
\end{equation}
This law is model independent. Expanding
$\Phi(x,\theta,\bar\theta)$ in powers of the spinor
(anticommuting, hence nilpotent) variables $\theta, \bar\theta$
produces a finite set of ordinary component fields $f(x)$, $\psi^\alpha (x), \ldots \,$.

${\cal N}$-extended supersymmetry can also be realized in the {\em chiral}
superspace ${\Bbb C}^{4\vert2N}$ which is {\em complex} and
involves only {\em half} of the spinor coordinates:
\begin{equation}\label{1.6}
\delta x^a_L =-2i\theta^i_L\sigma^a\bar\epsilon_i\;,
\hspace*{1.5cm}\delta
\theta^{\alpha}_{Li}=\epsilon^{\alpha}_{i}\;.
\end{equation}
The real superspace ${\Bbb R}^{4\vert4N}$ forms
a real hypersurface in the complex superspace ${\Bbb C}^{4\vert2N}$:
\begin{equation}
x^a_L=x^a+i\theta^i\sigma^a\bar\theta_i\;,\quad
\theta^{\alpha}_{Li} = \theta^\alpha\;. \label{1.7}
\end{equation}

The chiral superfields \cite{Ferr}
$\Phi(x_L,\theta_L)=\Phi(x+i\theta\sigma\bar\theta, \theta)$ defined in
${\Bbb C}^{4\vert2N}$ can be viewed as Grassmann analytic
superfields. Indeed, they obey the constraint
\begin{equation}\label{1.8}
\bar D_{\dot\alpha i} \Phi= 0\,,
\ee
where $\bar D_{\dot\alpha i}$ is the covariant (i.e., commuting
with the supersymmetry transformations) spinor derivative
\be
\bar D_{\dot\alpha i} = -{\partial\over\partial{\bar\theta}^{\dot\alpha i}}- i(\theta_{i}
\sigma^a)_{\dot\alpha}{\partial\over\partial x^a}\,.
\ee
Together with
\be
D^i_\alpha = {\partial\over\partial{\theta}^{\alpha}_ i}+ i(\sigma^a\bar\theta^{i}
)_{\alpha}{\partial\over\partial x^a}
\ee
they form an algebra similar to \p{N2}
\be
\{D^i_\alpha, \bar D_{\dot\alpha\,k}\} =
-2i\delta^i_k (\sigma^a)_{\alpha\dot\alpha}\frac{\partial}{\partial x^a}\,.\label{N2com}
\ee

In the basis $(x_L,\theta_L, \bar\theta)$ the derivative $\bar D_{\dot\alpha i}$
takes the ``short'' form $\bar D_{\dot\alpha i} = -{\partial/\partial{\bar
\theta}^{\dot\alpha i}}$. Then the constraint (\ref{1.8}) becomes
a sort of Grassmann Cauchy-Riemann condition
\begin{equation}\label{1.9}
{\partial\Phi\over\partial{\bar\theta}^{\dot\alpha}_{i}}=0
\end{equation}
which implies that $\Phi$ is a function of $\theta_L$ but not
of $\bar\theta$ (cf. the standard Cauchy-Riemann condition $\partial/\partial
\bar z f(z)= 0$ which means that the function $f$ depends on the variable
$z$ and not on its complex conjugate $\bar z$). This simple version of
Grassmann analyticity \cite{GA} works
effectively in ${\cal N}=1$ supersymmetry.

The important concept of Grassmann analyticity admits nontrivial generalizations
which underlie the ${\cal N}=2$ and ${\cal N}=3$
supersymmetric theories, and these generalized Grassmann analyticities constitute
the basis of the harmonic superspace approach.

In general, the fields appearing in the $\theta$-expansion of a superfield
form reducible supermultiplets. To single out the irreducible multiplets,
one should subject the carrier superfield to certain manifestly supersymmetric
constraints and/or admit some gauge freedom for it.

It should by emphasized that finding the adequate superspace for a
given theory is, as a rule, a nontrivial problem. The
superspaces ${\Bbb R}^{4\vert4N}$ and ${\Bbb C}^{4\vert2N}$ prove
to be appropriate for off-shell formulations only in the simplest
case of ${\cal N}=1$ supersymmetry. These ``standard'' superspaces are not so useful in the
extended (${\cal N}>1$) supersymmetric theories.

\subsection{Chirality as a key to ${\cal N}=1$ theories}

The chiral (${\cal N}=1$ analytic) superspace ${\Bbb C}^{4\vert2}$ forms the
basis of all ${\cal N}=1$ theories: they are either formulated
in terms of chiral superfields (matter and its self-couplings) or
follow from gauge principles which respect chirality (super Yang-Mills (SYM)
and supergravity (SG) theories and their couplings to matter).

The most general action of ${\cal N}=1$ matter is the action of n chiral
superfields $\Phi_A(x_L, \theta)\,$, $(A=1,\ldots n)\,$, and it is given by
\be
S_{\Phi} = \int d^4xd^4\theta K(\Phi_A, \bar\Phi_B) +
\left[ \int d^4 x_L V(\Phi_A) + \mbox{c.c} \right]. \lb{Phi}
\ee
In components, the first term gives a sigma model-type action, with the most general
$2n$-dimensional K\"ahler target metric for which $K(\phi, \bar\phi)$ is the K\"ahler
potential \cite{Zum}. The second term, after elimination of the auxiliary fields by their equations of motion,
yields the most general scalar potential of $\phi, \bar\phi$ consistent with ${\cal N}=1$ supersymmetry,
plus the appropriate Yukawa couplings of physical fermionic fields. Any other off-shell matter
representation of ${\cal N}=1$ supersymmetry (e.g., the so-called tensor ${\cal N}=1$ multiplet)
is described by the properly constrained ${\cal N}=1$ superfields related to the chiral ones via the
appropriate duality transformation \cite{N1Dual}.

The fundamental object (prepotential) of ${\cal N}=1$ SYM theory carrying the
irreducible field content of the off-shell ${\cal N}=1$ vector multiplet (gauge field
$b_m(x)$, gaugino $\psi_\alpha (x), \bar\psi_{\dot\alpha}(x)$ and the auxiliary
field $D(x)$, all taking values in the adjoint representation of some gauge group)
is the real scalar superfield $V(x,\theta, \bar\theta)\,$ \cite{SYM}. Its gauge transformation,
up to nonlinear terms, is given by \be
V'(x,\theta, \bar\theta) = V(x,\theta, \bar\theta) +
\frac{i}{2}\left(\Lambda(x_L, \theta) - \bar\Lambda(x_R, \bar\theta)\right) + {\cal O}(V) \,,
\lb{GaugeV} \ee where $\Lambda$ and $\bar\Lambda$ are conjugate
gauge-algebra valued superfield parameters, defined as unconstrained functions on the left- and
right-handed ${\cal N}=1$ chiral subspaces. The maximally reduced form of $V(x,\theta, \bar\theta)$
(Wess-Zumino gauge) is  as follows
\bea
&& V(x,\theta, \bar\theta) = \theta\sigma^n\bar\theta\,b_n +
(\bar\theta)^2\theta^\alpha \psi_\alpha +
(\theta)^2\bar\theta_{\dot\alpha}\bar\psi^{\dot\alpha}
+ (\theta)^2(\bar\theta)^2 D\,, \lb{WZ} \\
&&\delta A_n = \partial_n \lambda_0\,,\quad \lambda_0 \equiv
-\frac{1}{2}(\Lambda + \bar\Lambda)|_{\theta = \bar\theta = 0}\,. \nonumber
\eea
The fields in \p{WZ} are recognized as the irreducible off-shell ${\cal N}=1$
vector multiplet.
From \p{GaugeV} it follows that the fundamental
gauge group of ${\cal N}=1$ SYM theory is represented by chiral superfield gauge parameters.
The differential geometry constraints defining this theory in the superspace ${\Bbb R}^{4\vert4}$
were given in \cite{N2SYM} \be \{{\cal D}_\alpha, {\cal D}_\beta\} = \{\bar{\cal
D}_{\dot\alpha}, \bar{\cal D}_{\dot\beta}\} = 0\,.  \lb{N1symC}
\ee
Here ${\cal D}_\alpha = D_\alpha + i {\cal A}_\alpha$ is a gauge-covariantized
spinor derivative. These constraints are just the integrability conditions for
the existence of chiral ${\cal N}=1$ superfields in the full interacting case, thus
expressing the fact that ${\cal N}=1$ SYM theory is fully determined by the requirement of
preservation of ${\cal N}=1$ chirality, the simplest form of Grassmann analyticity.

Finally, the superspace geometry of ${\cal N}=1$ SG \cite{N1SG} is also fully
fixed by the chirality-preservation principle.

The underlying gauge group of conformal ${\cal N}=1$ SG is just the
group of general diffeomorphisms of the chiral superspace \cite{OS}: \be \delta x^m_L =
\lambda^m(x_L, \theta_L)\,, \quad \delta \theta^\mu_L = \lambda^\mu(x_L,
\theta_L)\,,\lb{ChirDif} \ee with $\lambda^m, \lambda^\mu$ being arbitrary
complex functions of their arguments.
The basic gauge prepotential of conformal ${\cal N}=1$ SG is an axial-vector superfield $H^m(x,\theta,
\bar\theta)$ appearing as the imaginary part of the bosonic chiral coordinate,
\be
x_L^m = x^m + i H^m(x,\theta, \bar\theta)\,, \quad \theta^\mu = \theta_L^\mu\,, \quad  \bar\theta^{\dot\mu}
= \overline{(\theta^\mu)}.\lb{Emb1}
\ee
It possesses a nice geometric meaning: it specifies the superembedding of
real ${\cal N}=1$ superspace $(x^m, \theta^\mu, \bar\theta^{\dot\mu})$
as a hypersurface into the complex chiral ${\cal N}=1$ superspace $(x_L^m,
\theta^\mu_L)\,$\footnote{It was shown in \cite{I11} that the geometric meaning of the ${\cal N}=1$ SYM
prepotential $V(x,\theta, \bar\theta)$ is to some extent similar to that
of $H^m(x,\theta, \bar\theta)$. The superfield $V$ also specifies a real $(4|4)$
dimensional hypersurface, this time in the product of ${\cal N}=1$ chiral superspace
and the internal coset space $G^c/G$, where $G^c$ is the complexification of the gauge group $G\,$.}.  Through the relations
\p{Emb1}, the transformations \p{ChirDif} generate field-dependent nonlinear
transformations of the ${\cal N}=1$ superspace coordinates $(x^m, \theta^\mu,
\bar\theta^{\dot\mu})$ and of the superfield $H^m(x, \theta, \bar\theta)\,$.
The field content of $H^m$ can be revealed in the WZ gauge:
\be H^m_{WZ} = \theta\sigma^a\bar\theta\,e^m_a + (\bar\theta)^2
\theta^\mu \psi^m_\mu + (\theta)^2 \bar\theta_{\dot\mu}\bar\psi^{m\dot\mu} +
(\theta)^2(\bar\theta)^2 A^m\,. \label{WZH}
\ee
Here one finds the vierbein $e^m_a$ representing the conformal graviton (gauge-independent spin 2 off-shell),
the gravitino $\psi^m_\mu$ (spins $3/2$), and the gauge field $A^m$ (spin 1)
of the local $\gamma_5$ R-symmetry, just $(8 +8)$ off-shell degrees of freedom
forming
${\cal N}=1$ Weyl multiplet. Various  versions of
 ${\cal N}=1$ Einstein SG actions can be constructed as the actions of various ${\cal N}=1$ superfield compensators
in the background of ${\cal N}=1$ Weyl multiplet. All of these actions are related to
the version with a chiral compensator via appropriate duality transformations. Once again,
the basic differential supergeometry constraints of ${\cal N}=1$ SG have the interpretation
of  integrability conditions for the existence of ${\cal N}=1$ chiral superfields in the full
curved case.

In theories with extended ${\cal N}=2$ and ${\cal N}=3$ supersymmetries this remarkable
chirality-preservation principle is substituted by the principle
of preservation of a different type of Grassmann analyticity, the Grassmann
harmonic one.
\setcounter{equation}{0}

\section{${\cal N}=2$ Harmonic superspace}

\subsection{Why standard superspaces are not enough for ${\cal N}=2$ case}

In the framework of the standard superspaces ${\Bbb R}^{4\vert8}$
and ${\Bbb C}^{4\vert4}$ it proved
impossible to find an off-shell action principle for an
unconstrained description of all ${\cal N}=2$ supersymmetric
theories.

The basic problem with extended superspace
$(x^m, \theta_{i}^\alpha, \bar\theta^{\dot\alpha i})$ was that the
corresponding superfields, due to the large number of Grassmann coordinates,
contain too many irreducible supermultiplets. So they should be either strongly
constrained or subjected to gauge transformations with an {\it a priori} unclear
geometric meaning. Another problem was that some constraints imply the
equations of motion for the fields involved, which makes impossible to find
an invariant off-shell action for them. For instance, in the ${\cal N}=2$ case  the simplest matter
multiplet (the analog of ${\cal N}=1$ chiral multiplet) is the hypermultiplet \cite{FS, CentrCharg1} which is
represented by a complex $SU(2)$ doublet superfield $q^{i}(x, \theta,
\bar\theta)$ ($i=1,2$) subjected to the constraints \be D^{(i}_\alpha q^{k)} = \bar
D^{(i}_{\dot\alpha} q^{k)} = 0\,. \lb{HypC} \ee Here
$D^i_\alpha, \bar D^k_{\dot\alpha}$ are ${\cal N}=2$ spinor
covariant derivatives satisfying the relations \p{N2com}.
On shell this supermultiplet contains four scalar fields forming an
$SU(2)_A$ doublet $f^i(x)$ and two isosinglet spinor fields
$\psi^\alpha(x), \bar \kappa^{\dot\alpha}(x)$.
Using \p{N2com}, it is
a direct exercise to check that \p{HypC} gives rise to the equations of motion
for the physical component fields in $q^i = f^i + \theta^{i\alpha}\psi_\alpha +
\bar\theta^i_{\dot\alpha}\bar\chi^{\dot\alpha} + \ldots\,$, viz.,
\bea
\Box f^i
= 0\,, \quad \partial_m\psi \sigma^m = \sigma^m\partial_m\bar\chi =
0\,.\lb{MassHyp}
\eea
This phenomenon is a reflection of the ``no-go'' theorem
\cite{nogo,6D} stating that no off-shell representation for a hypermultiplet in its
``complex form'' ({\it i.e.}, with the bosonic fields arranged in an $SU(2)$ doublet) can
be achieved with any {\it finite} number of auxiliary fields. At the time it was not clear
if there existed a way to circumvent this theorem and to
write some kind of an off-shell action for the hypermultiplet.

A further problem was the lack of a geometric unconstrained formulation
of ${\cal N}=2$ SYM theory, similar to the prepotential formulation of ${\cal N}=1$ SYM.
The differential geometry constraints defining this theory were given in
\cite{N2SYM} \be \{{\cal D}^{(i}_\alpha, {\cal D}^{k)}_\beta\} = \{\bar{\cal
D}^{(k}_{\dot\alpha}, \bar{\cal D}^{i)}_{\dot\beta}\} = \{{\cal
D}^{(i}_{\alpha}, \bar{\cal D}^{k)}_{\dot\beta}\} = 0\,,  \lb{N2symC} \ee where
${\cal D}_\alpha^i = D^i_\alpha + i {\cal A}^i_\alpha$ is a gauge-covariantized
spinor derivative. Mezincescu was the first to find the solution of these
constraints in the Abelian case through an unconstrained prepotential
\cite{LMe}. However the latter possesses a non-standard dimension -2, and the
corresponding gauge freedom does not admit a geometric interpretation. It was unclear
whether something like the nice  geometric interpretation of the
${\cal N}=1$ SYM gauge group and prepotential $V$ could be found in the ${\cal N}=2$ (and ${\cal N}>2$) case.
The same problem existed for ${\cal N}=2$ superfield SG.

In \cite{GA} Galperin, Ivanov and Ogievetsky observed that extended supersymmetry, besides
standard chiral superspaces generalizing the ${\cal N}=1$ one, also admits some other
type of invariant subspaces which were called ``Grassmann-analytic''. Like in the
case of chiral superspaces, these subspaces are obtained by passing to some new
basis in the general superspace, such that the spinor covariant derivatives with
respect to a subset of the Grassmann variables become ``short'' in it. Then
one can impose Grassmann Cauchy-Riemann conditions with respect to these
variables, while preserving the full extended supersymmetry. In the simplest ${\cal N}=2$ case, allowing the $U(2)$
automorphism symmetry to be broken down to $O(2)$, and making the appropriate
shift of $x^m$, one can define the complex ``$O(2)$ analytic subspace'' \be
(\tilde{x}^m\,, \;\theta^1_\alpha + i\theta^2_\alpha\,,
\;\bar\theta^1_{\dot\alpha} + i\bar\theta^2_{\dot\alpha})\,, \lb{O2Anal} \ee
which is closed under ${\cal N}=2$ supersymmetry, and the related Grassmann-analytic
superfields. It was natural to assume that this new type of analyticity plays a
fundamental role in extended supersymmetry, similar to chirality in the ${\cal N}=1$ case. In
\cite{Anat} it was found that the hypermultiplet constraints \p{HypC} imply that
different components of ${\cal N}=2$ superfield $q^i$ ``live'' on different
$O(2)$-analytic subspaces. Since \p{HypC} is $SU(2)$ covariant, it was tempting
to ``$SU(2)$- covariantize'' the $O(2)$ analyticity.

All these problems were solved with the invention of harmonic superspace
\cite{aG5}-\cite{aG8}, \cite{Book}. The papers \cite{OS}, \cite{I11}, \cite{GA} and \cite{Anat}
turned out to be the basic benchmarks on the way towards this new concept.

\subsection{Harmonic superspace: the definition}
${\cal N}=2$ harmonic superspace (HSS) ${\Bbb H}^{4+2\vert8}$
is defined as the product
\be
{\Bbb H}^{4+2\vert8} = {\Bbb R}^{4\vert8}\otimes S^2 = (x^m, \theta_{\alpha\;i},
\bar\theta_{\dot\beta}^k )\otimes (u^+_i, u^-_k)\,. \lb{HSS}
\ee
Here $S^2 \sim SU(2)_A/U(1)$, with $SU(2)_A$ being
the automorphism group of the ${\cal N}=2$ superalgebra.
The internal 2-sphere $S^2$ is realized in
a parametrization-independent way by the lowest (isospinor) $SU(2)_A$
harmonics
\be
S^2 \in (u^+_i, u^-_k), \quad u^{+i}u_i^- =1, \quad u^{\pm}_i \rightarrow
\mbox{e}^{\pm i\lambda}u^{\pm}_i~.
\ee
It is assumed that nothing depends on the $U(1)$ phase $\mbox{e}^{i\lambda}$,
so one effectively deals with the 2-sphere $S^2 \sim SU(2)_A/U(1)$.
The superfields given on \p{HSS} (harmonic ${\cal N}=2$ superfields)
are assumed to be expandable in harmonic series on $S^2$,
with the set of all symmetrized products of $u^+_i, u^-_i$
as a basis. These series are fully specified by the $U(1)$ charge
of the given superfield.

The main advantage of HSS is the existence of an invariant subspace in it, the
${\cal N}=2$ analytic HSS with half of the original odd coordinates \bea
&&\HA^{(4+2\vert 4)} =
\left(x^m_A, \theta^{+\alpha}_A, \bar\theta^{+\dot\alpha}_A,
u^{\pm i}\right)  \equiv \left(\zeta, u^{\pm}_i  \right)~, \label{22} \\
&& x^m_A = x^m -2i\theta^{(i}\sigma^m\bar\theta^{k)}u^+_iu^-_k~,
\quad \theta^{+ \alpha} = \theta^{\alpha\,i}u^{+}_i~, \;
\bar\theta^{+b\dot\alpha} = \bar\theta^{\dot\alpha\,i}u^{+}_i~. \label{222}
\eea
This is just the $SU(2)$ covariantization of the $O(2)$ analytic superspace \p{O2Anal}.
It is closed under ${\cal N}=2$ supersymmetry transformations and is real with respect to
the special involution which is the product of the ordinary complex
conjugation and the antipodal map (Weyl reflection) of $S^2$. As will be shown later, all ${\cal N}=2$
supersymmetric theories admit off-shell formulations in terms of unconstrained
superfields defined on \p{22}, the {\it Grassmann analytic}
${\cal N}=2$ superfields. But before passing to this issue, let us briefly
describe the basic technical tools of the harmonic superspace approach.

\subsection{Harmonic calculus on $S^2$}
\vspace{0.3cm}

In the HSS approach, the harmonic sphere $S^2 \sim SU(2)/U(1)$
is coordinatized by ``zweibeins'' $u^{+
i}, u^{-}_i=\overline{u^{+i}}$ having $SU(2)$ indices $i$ and
$U(1)$ charges $\pm$. The solution of the constraint
\begin{equation}
u^{+i} u_i^{-}=1\;, \label{1.23}
\end{equation}
is the matrix
\begin{equation}
\parallel u\parallel  = \left(\begin{array}{cc} u^+_1 & u^-_1 \\
u^+_2 & u^-_2 \end{array} \right) = {1 \over \sqrt{1+t\bar t}}
\left(\begin{array}{cc} e^{i\psi} & -\bar t e^{-i\psi} \\
te^{i\psi} & e^{-i\psi}\end{array} \right), \ \ \ 0\leq \psi <
2\pi \label{1.24}
\end{equation}
corresponding to the group $SU(2)$ in the stereographic
parametrization $(t,\bar t, \psi)\,$ (one can equally choose any other
specific parametrization). To realize the coset space
$S^2 \sim SU(2)/U(1)$, the zweibeins have to be defined up
to a $U(1)$ phase corresponding to a transformation of the $U(1)$
group in the coset denominator:
\begin{equation}
u^{+'}_i = e^{i\alpha}u^{+}_i\;, \quad u^{-'}_i \label{1.25}
=e^{-i\alpha}u^{-}_i\;.
\end{equation}
So, the phase $\psi$ in the parametrization \p{1.24} is inessential
and one effectively deals only with the complex coordinates
$t,\bar t$. For the phase not to show up at all, the
``functions'' on $S^2$ must possess a {\em definite} $U(1)$
charge $q$ and, as a consequence, all the terms in their harmonic
expansion must contain only products of zweibeins $u^{+}\;,u^{-}$
of the given charge $q$. For instance, for $q=+1$
\begin{equation}
f^{+}(u)=f^iu^{+}_i + f^{(ijk) }u_{(i}^{+}u_j^{+}u_{k)}^{-} +
\ldots\;. \label{1.26}\end{equation} Such quantities undergo
homogeneous $U(1)$ phase transformations, according to their
overall charge. This restriction on the harmonic functions is
called $U(1)$ charge preservation. In each term in \p{1.26}
a complete symmetrization of the indices $i,j, k,\ldots$ is assumed. Indeed, any product of the harmonics with a fixed
overall $U(1)$ charge can be reduced to their symmetrized product plus lower-rank symmetrized products, using the completeness relation
\begin{equation}
u^+_iu^-_k - u^+_k u^-_i = \varepsilon_{ik} \label{ComplS}
\end{equation}
following from the basic constraint \p{1.23}.

In fact, the zweibeins $u^+_i,u^-_i$ are the fundamental spin
$1/2$ spherical harmonics, and
(\ref{1.26}) is an example of a harmonic decomposition on $S^2$.
This is why $u^+_i,u^-_i$ are referred to as {\em harmonic
variables} (or simply ``harmonics'').

It is instructive to list the following specific features of the
treatment of the $S^2$ expansion in the harmonic space approach
as compared to the standard textbooks and reviews
on harmonic analysis (e.g. in \cite{aV1,whit}).
\begin{itemize}
\item The harmonics themselves are regarded as the $S^2$ coordinates.
This allows one to avoid using any explicit parametrization
like the stereographic one \p{1.24}. What really matters is
the defining constraint \p{1.23} together with the requirement of
$U(1)$ charge preservation. If one exploits
the harmonics $u^+_i,u^-_i$ as ``global'' coordinates on $S^{2}$,
there arises no need in several  ``charts'' to cover the sphere
(which is inevitable when using any explicit parametrization).
If one has succeeded in solving some equation in terms of
harmonics, then the solution obtained is well defined on the
entire sphere.
\item The harmonic expansions go over symmetrized products of harmonics instead of sets
of special functions, like the Jacobi polynomials or the spherical
functions. As a result, the coefficients in the harmonic expansions (like $f^i,
f^{(ijk)}\;,$ ... in \p{1.26}) transform as irreducible
representations of the $SU(2)$ group of the coset numerator. This
is of special value in ${\cal N}=2$ supersymmetry because the ${\cal N}=2$
supermultiplets are classified according to the $SU(2)$ automorphism group.
\end{itemize}

In accord with the treatment of $u^{\pm}_i$ as the $S^2$ coordinates, one may introduce
two covariant derivatives compatible with the constraint \p{1.23} and having $U(1)$ charges $+2$ and $-2$:
\begin{equation}
D^{++}=u^{+i}{\partial\over {\partial{u^{-i}}}}\quad \mbox{and}
\quad D^{--}=u^{-i}{\partial\over {\partial{u^{+i}}}}\;.
\label{1.27}
\end{equation}
They play a major role in the harmonic superspace approach and are referred to as
harmonic derivatives\footnote{This essential use of the derivatives with respect to
the additional bosonic co-ordinates is the characteristic
feature of the HSS approach as compared, e.g., with the superfield approach based on the
projective ${\cal N}=2$ superspace \cite{LiRo}. As argued in \cite{Kuz}, the latter formalism
is a particular version of the HSS one corresponding to a special parametrization of
the harmonics.}. These derivatives commute with the original $SU(2)_A$
group and form, in their own, an $SU(2)$ algebra:
\be
\left[D^{++}, D^{--} \right] = D^0 \equiv u^{+i}{\partial\over {\partial{u^{+i}}}}
- u^{-i}{\partial\over {\partial{u^{-i}}}}\,, \quad
\left[D^0, D^{\pm\pm}\right] = \pm 2 D^{\pm\pm}\,,
\ee
where the operator $D^0$ also commutes with $SU(2)_A$ and represents the third
covariant differential operator on $S^3 \sim SU(2)_A\,$. When acting on functions on $SU(2)_A/U(1)$
like $f^{(q)}$ in \p{1.26}, it takes the fixed value,
\be
D^0 f^{(q)}(u) = q f^{(q)}(u)\,,
\ee
{\it i.e.}, it just counts the harmonic $U(1)$ charge. Thus the covariant derivations on the coset $S^2$ are defined
by the operators $D^{\pm\pm}$, in accordance with the dimension of this coset.
{}From the definition \p{1.27} follow the obvious rules for the
action of $D^{\pm\pm}$ on the harmonics
\begin{equation}
D^{\pm\pm}u^\pm_i=0\;,\qquad D^{\pm\pm}u^\mp_i= u^\pm_i\;.
\label{1.29}\end{equation}

To get a better feeling how convenient it is to use $u^{\pm}_i$ as the coordinates of $S^2$,
let us consider the simple harmonic differential equation
\begin{equation}
D^{++}f^+(u)=0 \;. \label{1.30}
\end{equation}

\noindent In harmonics its solution is immediately obtained from
(\ref{1.26}):
\begin{equation}
f^+=f^i u^+_i \label{1.31}\end{equation} where $f^i$ are arbitrary
constants. Indeed, $f^+$ has this form because all other terms in
its harmonic expansion include $u^-$. For the general harmonic function $f^{(q)}$ with $q \geq 0$ the solution of the equation
analogous to \p{1.30} is given by
\be
f^{(q)}(u) = f^{(i_1i_2 \cdots i_q)}u^{+}_{i_1}\cdots u^+_{i_q}\,,
\ee
where the symmetric tensor $f^{(i_1i_2 \cdots i_q)}$ represents
an irreducible $SU(2)_A$ multiplet with  isospin $q/2$.
Another important property is
\be
D^{++} f^{(q)}( u) = 0 \; \Rightarrow \; f^{(q)} = 0\,, \quad \mbox{iff} \;\; q< 0\,.
\ee
It can be easily proved using the harmonic expansions \p{1.26} and the property \p{1.29}.

Finally, in order to be able to construct invariant actions
one needs to define an integration on the two-sphere $S^2$. In the
harmonic approach it is introduced by the following formal rules:
\begin{equation}
\int du\; 1=1\;,\qquad \int du\;  u^+_{(i_1}..u^+_{i_k}
u^-_{i_{k+1}}.. u^-_{i_{k+l})}=0\;. \label{1.34}\end{equation}
This definition means the vanishing of the integrals of any
spherical function with non-zero isospin (represented by symmetrized products
of harmonics). It admits integration by parts, etc.
These rules can be justified by the use of some specific
parametrization for the harmonics, e.g., \p{1.24}. In this parametrization,
the same harmonic integral is given as
\begin{equation}
\int du\;  f^{(q)}(u) \equiv {i\over 4\pi^2} \int^{2\pi}_0 d\psi \int {dt\wedge d\bar
t\over (1+t\bar t)^2} f^{(q)}(t,\bar t,\psi)\;.
\label{4.2.2}\end{equation}
However, the abstract form of the $u$-integral defined by the rules \p{1.34}  is most convenient in a field theory.

\subsection{Grassmann harmonic analyticity}
The analytic basis in the harmonic ${\cal N}=2$ superspace ${\Bbb H}^{4+2\vert8}$ is defined as the following
set of coordinates
\bea
\underline{\mbox{Analytic \;basis}}: \quad \left(x^m_A, \theta^{+ \alpha}, \bar\theta^{+ \dot\alpha},
u^{\pm i}, \theta^{-\alpha}, \bar\theta^{-\dot\alpha}\right) := \left(\zeta,, u, \theta^-, \bar\theta^-\right), \label{AnBas}
\eea
where $x^m_A, \theta^{+ \alpha}, \bar\theta^{+ \dot\alpha}$ were defined in \p{222} and
$\theta^{-\alpha} = \theta^{\alpha \,i} u^-_i\,$,
$\bar\theta^{-\dot\alpha} = \bar\theta^{\dot\alpha\,i}u^-_i\,$. The original parametrization \p{HSS}
is referred to as the ``$\underline{\mbox{Central basis}}$''. The main feature of the analytic basis is that it
makes manifest the existence of the harmonic analytic subspace \p{22} closed under the ${\cal N}=2$
supersymmetry transformations. Correspondingly, defining the harmonic projections of the spinor
covariant derivatives
\be
D^\pm_\alpha = D^i_\alpha u^\pm_i\,, \quad \bar D^\pm_{\dot\alpha} = \bar D^i_{\dot\alpha}u^\pm_i\,,\lb{DefDpm}
\ee
it is straightforward to find that the derivatives $D^+_\alpha, \bar D^+_{\dot\alpha}$ become ``short''
in the analytic basis
\be
D^+_\alpha = \frac{\partial}{\partial \theta^{- \alpha}} := \partial^+_{\alpha}\,, \quad
\bar D^+_{\dot\alpha} = \frac{\partial}{\partial \bar\theta^{- \dot\alpha}} := \partial^+_{\dot\alpha}\,,
\ee
like, e.g.,  the ${\cal N}=1$ derivative $\bar D_{\dot\alpha}$ in the left-chiral basis (cf.
\p{1.8}, \p{1.9}). Now, consider a superfield on ${\Bbb H}^{4+2\vert8}$,
$\Phi^{(q)} (x, \theta, \bar\theta, u)\,$, where $q$ is the external harmonic $U(1)$ charge,
and impose on it the manifestly supersymmetric conditions
\be
D^+_\alpha \Phi^{(q)} = 0\,, \quad \bar D^+_{\dot\alpha} \Phi^{(q)} = 0\,. \label{AnalCondi}
\ee
This set of constraints is self-consistent in view of the obvious integrability conditions
\be
\{D^+_\alpha, D^+_\beta \} = \{\bar D^+_{\dot\alpha}, \bar D^+_{\dot\beta} \} =
\{D^+_\alpha, \bar D^+_{\dot\beta} \} = 0\,. \label{IntDD}
\ee
Since in the analytic basis the derivatives $D^+_\alpha, \bar D^+_{\dot\alpha}$ are reduced to
the partial ones, the conditions \p{AnalCondi} become none other than one more
example of the Grassmann Cauchy-Riemann conditions: they mean that in this basis the harmonic superfield $\Phi^{(q)}$
does not depend on half of the Grassmann coordinates, viz. $\theta^-_\alpha,
\bar\theta^-_{\dot\alpha}\,$:
\be
\p{AnalCondi} \;\; \Rightarrow \;\; \Phi^{(q)}(x, \theta, \bar\theta, u^\pm) =
\varphi^{(q)} (\zeta, u^\pm)\,,
\ee
and so ``lives'' on the analytic subspace $\HA^{(4+2\vert 4)}$ defined in \p{22}.
The superfields $\varphi^{(q)}(\zeta^A, u^\pm)$ are called ``harmonic analytic superfields''.
This type of Grassmann analyticity clearly generalizes ${\cal N}=1$ chirality (cf. \p{1.8} and \p{1.9})
and is called ``Grassmann harmonic analyticity''. All ${\cal N}=2$ matter, SYM and supergravity theories
have adequate off-shell description in terms of the appropriate analytic ${\cal N}=2$ superfields.

Since the analytic harmonic coordinates $\theta^+_\alpha, \bar\theta^+_{\dot\alpha}$ carry
the harmonic $U(1)$ charge $+1$, the coefficients in the $\theta$ expansion of
$\varphi^{(q)}(x_A, \theta^+, \bar\theta^+, u)$ have $U(1)$ charges ranging from $q$ to
$q-4$, so that the total $U(1)$ charge is always equal to $q$:
\be
\varphi^{(q)} = f^{(q)}(x, u) + \theta^{+\alpha}\psi^{(q-1)}(x, u) +
\bar\theta^+_{\dot\alpha}\chi^{(q-1)\dot\alpha} + \ldots\;. \label{expanAnal}
\ee
All these component fields are assumed to be expandable in  harmonic series of
the type \p{1.26} on $S^2 \sim SU(2)_A/U(1)\,$. A very important and surprising
property follows from this: the analytic harmonic superfields contain {\it infinitely} many fields
which can be assembled into infinite series of irreducible supermultiplets with the same fixed
superspin $Y$ and increasing superisospins $I$ (with values $I = \vert \frac{q}{2} -1\vert + n\,$,
$n = 0, 1, 2, \ldots $)\footnote{The superspin is the analog of the Poincar\'e spin. The superisospin of a given supermultiplet
coincides with the isospin of the state with the highest spin (see, e.g., \cite{Book}).}. In some cases (${\cal N}=2$ matter) these infinite ``tails'' of fields become
auxiliary while in other cases (${\cal N}=2$ SYM and ${\cal N}=2$ SG) they are pure gauge.

The harmonic derivative $D^{++}$ commutes with the spinor derivatives $D^+_\alpha, \bar D^+_{\dot\alpha}\,$,
\be
[D^{++}, D^+_\alpha] = [D^{++}, \bar D^+_{\bar\alpha}] = 0\,, \label{D++D}
\ee
and so it preserves harmonic analyticity: acting on $\varphi^{(q)}$, it again yields an analytic
superfield. In the analytic basis it takes the form
\bea
&& D^{++}_A = \partial^{++} - 2i (\theta^+\sigma^m \bar\theta^+) \partial_m + \theta^{+\alpha}{\partial^{+}_{\alpha}} + \bar\theta^{+\dot\alpha}\bar{\partial}^{+}_{\dot\alpha} \,,
\nonumber \\
&& \partial^{++} :=
u^{+ i}\frac{\partial}{\partial u^{- i}}\,. \lb{D++Anal}
\eea
The $U(1)$-charge counting operator $D^0$ obviously preserves  harmonic analyticity too, in the analytic basis it reads
\bea
&& D^0_A = \partial^{0} + \theta^{+\alpha}{\partial^{-}_{\alpha}} + \bar\theta^{+\dot\alpha}\bar{\partial}^{-}_{ \dot\alpha}
- \theta^{-\alpha}{\partial^{+}_{\alpha}} -\bar\theta^{-\dot\alpha}\bar{\partial}^{+}_{\dot\alpha}\,, \nonumber \\
&& \partial^{0} :=
u^{+ i}\frac{\partial}{\partial u^{+ i}} - u^{- i}\frac{\partial}{\partial u^{-i}}.\label{D0Anal}
\eea

\setcounter{equation}{0}

\section{${\cal N}=2$ matter and gauge theories in harmonic superspace}

Now we will overview the formulations, main
ideas and results related to ${\cal N}=2$ matter and SYM theories in
the harmonic superspace approach as a pre-requisite to the analogous formulations of
${\cal N}=2$ supergravities.

\subsection{${\cal N}=2$ matter hypermultiplets}
\label{hypt}
One of the main results obtained within the HSS approach is the discovery of an off-shell
formulation of the Fayet-Sohnius ${\cal N}=2$ hypermultiplet, thus circumventing the no-go theorems
which do not seem to allow such a formulation \cite{nogo,6D}. After the HSS formulation
has been found, it became clear  that the loophole of these theorems was the implicit assumption about
the {\it finite} number of admissible auxiliary fields. The basic feature of the off-shell
HSS description of the hypermultiplet is the {\it infinite} set of auxiliary fields.

With the help of the harmonics $u^\pm_i$ the constraints \p{HypC} can be given another,
more suggestive form. Namely, introducing a general ${\Bbb H }^{4+2\vert8}$ superfield
$q^+$
\be
q^+(x, \theta, \bar\theta, u)  = q^i(x, \theta, \bar\theta) u^+_i +
q^{(ikl)}(x,\theta, \bar\theta)u^+_{(i}u^+_ku^-_{l)} + \ldots\,,
\ee
one can equivalently rewrite \p{HypC} as
\be
\mbox{(a)} \; D^+_\alpha q^{+} = \bar D^+_{\dot\alpha} q^+ = 0\,, \qquad \mbox{(b)}\; D^{++} q^+ = 0\,,
\lb{Constrq}
\ee
where $D^+_\alpha = D^i_\alpha u^+_i, \bar D^+_{\dot\alpha} = \bar D^i_{\dot\alpha} u^+_i$
(recall \p{DefDpm}). Indeed, (\ref{Constrq}b) implies that $q^+ = q^iu^+_i$, in the same way as
\p{1.30} implies \p{1.31}, then (\ref{Constrq}a) gives just \p{HypC}:
\be
D^iq^ku^+_i u^+_k = \bar D^i_{\dot\alpha}q^k u^+_iu^+_k = 0 \quad \Rightarrow \quad D^{(i}q^{k)} =
\bar D^{(i}_{\dot\alpha}q^{k)} = 0
\ee
(one can take off  the symmetric product $u^+_i u^+_k$ in view of the arbitrariness of the harmonics).
The HSS constraints \p{Constrq} are self-consistent, since the differential operators in them
satisfy the integrability conditions \p{IntDD}, \p{D++D}.

The advantage of rewriting \p{HypC} in the form \p{Constrq} is revealed in the analytic basis.
As explained in Sec 3.4, in this basis eqs. (\ref{Constrq}a) are Grassmann analytic Cauchy-Riemann
conditions stating that $q^+$ is the analytic harmonic ${\cal N}=2$ superfield
\be
\mbox{(\ref{Constrq}a)} \quad \Rightarrow \quad q^+ = q^+(\zeta,  u^\pm)\,.
\ee
These are purely kinematical conditions, like the ${\cal N}=1$ chirality condition. All the dynamical aspects of
\p{HypC} now prove to be concentrated in (\ref{Constrq}b)
\be
D^{++}_A q^+(\zeta, u) = \left(\partial^{++}
- 2i \theta^+\sigma^m\bar\theta^+\partial_m\right) q^+ (\zeta, u) = 0 \label{DeqM}
\ee
(recall the analytic basis form \p{D++Anal} of $D^{++}$). It can be easily checked that
this equation eliminates all the infinite sets of fields present in
the $S^2$ harmonic expansions of the component fields and gives rise just to the equations of motion
\p{MassHyp} for the physical components. The most striking point is that \p{DeqM} can be derived from
the invariant off-shell action
\be
S_q^{free} = -\int dud\zeta^{(-4)}\; \tilde q^+ D^{++} q^+\,.
\label{ActqFr}
\end{equation}
Here, the operation $\tilde{}$ is a special involution preserving the
analytic harmonic superspace \p{22} (it is reduced to ordinary
complex conjugation for the $u$-independent quantities), and the
integration measure of the analytic superspace is defined as
\begin{equation}
d\zeta^{(-4)}=d^4xd^2\theta^+d^2\bar\theta^+\;. \label{1.47}
\end{equation}
This measure carries negative $U(1)$ charge because Grassmann
integration is equivalent to differentiation with
respect to the odd coordinates of the analytic superspace
$\theta^+_\alpha, \bar\theta^+_{\dot\alpha}$.

It is worth pointing out once more that the analytic superfield $q^+$ is
unconstrained in the off-shell action \p{ActqFr}, and its harmonic
expansion contains an {\em infinite number} of auxiliary
fields. This is how HSS manages to circumvent the
no-go theorem \cite{nogo,6D}.

Now it is rather straightforward to generalize \p{ActqFr} by including
general self-interactions. One introduces $n$ hypermultiplet
superfields $q^{+}_{a} (\zeta, u)$ ($\overline{(q^{+}_{a})} =
\Omega^{ab}q^{+}_{b}~,\;$ $\Omega^{ab} = -\Omega^{ba}$; $a,b = 1, \dots 2n$ )
and writes the following general off-shell action:
\be  \label{333}
S_q = \int du
d\zeta^{(-4)} \left\{q^{+}_{a}D^{++}q^{+a} + L^{+4}(q^+, u^+, u^-) \right\}~.
\ee
Here the indices $a, b$ are raised and
lowered by the $Sp(n)$ skew-symmetric tensors $\Omega^{ab},
\Omega_{ab}$, $\Omega^{ab}\Omega_{bc} = \delta^a_c $.
The interaction Lagrangian $L^{+4}$ is an arbitrary function of its arguments, the only
restriction is its harmonic $U(1)$ charge $+4$ needed to balance that of the superspace measure.
After eliminating the infinite sets of auxiliary fields by their (now nonlinear)
equations of motion, one gets the most general self-interaction of $n$
hypermultiplets. In the bosonic sector it yields a generic sigma model with
a $4n$-dimensional hyper-K\"ahler (HK) target manifold in accord with the theorem
of Alvarez-Gaum\'e and Freedman about the one-to-one correspondence between
${\cal N}=2$ supersymmetric sigma models and HK manifolds \cite{AGF} \footnote{HK manifolds are
$4n$-dimensional Riemannian manifolds which admit a triplet of
covariantly constant complex structures forming the algebra of
quaternionic units or, equivalently, such that their holonomy group
lies in $Sp(n)$. In the HSS approach, the triplet of complex structures is
parametrized by harmonics \cite{HKpaper}.}. In general, the
action \p{333} and the corresponding HK sigma model possess no isometries.
The action \p{333} is the ${\cal N}=2$ analog of the general ${\cal N}=1$ matter action \p{Phi},
the object $L^{+4}$ being the HK potential, analog of the K\"ahler potential $K(\Phi, \bar\Phi)$
of ${\cal N}=1$ supersymmetric sigma models. It encodes the complete
information about the local properties of a given HK manifold.
Taking some specific $L^{+4}$, one
gets the explicit form of the relevant HK metric after eliminating the auxiliary
fields from \p{333}. So, the general hypermultiplet action \p{333} provides an
efficient universal tool for {\it explicit} construction of   HK metrics.
For instance, the well-known 4-dimensional HK Taub-NUT metric corresponds to the
choice $L^{+4} \sim (\tilde q^+ q^+)^2$ \cite{TNpaper, Book}. Other HK metrics of this kind were constructed in analogous way in \cite{GIOT,GiVa}.
It is also easy to find, e.g.,  $L^{+4}$
yielding the well-known general Gibbons-Hawking Ansatz \cite{GH} for the 4-dimensional HK metrics
with at least one tri-holomorphic isometry (i.e. the one commuting with supersymmetry)
\cite{HKpaper, Book}.

Following the general recipe of ref. \cite{AFpot},  potential terms (including possible mass terms)
can be introduced for any action having at least one tri-holomorphic isometry by introducing a central
charge which is identified with the isometry generator times a mass-like parameter
({\it i.e.} via a mechanism \`a la Scherk-Schwarz \cite{ShSch}).
Specifically, this is achieved by extending the harmonic derivative \p{D++Anal} in \p{333} or \p{ActqFr}
as \cite{Book}
\be
D^{++} \rightarrow D^{++} + i [(\theta^+)^2 - (\bar \theta)^2]\frac{\partial}{\partial x^5}\,,\lb{ExtenD}
\ee
where $\partial/\partial x^5$ is the central charge, and choosing
\be
\frac{\partial}{\partial x^5} q^{+ a} = m\lambda^{+ a}(q, u)\,,
\ee
where $\lambda^{+ a}(q, u)$ is the Killing vector of the isometry.
The form of the scalar potential is completely determined by the Killing vector and the form of
the corresponding HK metric. In particular, in the free case ($L^{+4} = 0$),
the only possible effect of the extension \p{ExtenD} is the appearance of mass terms for the
physical fields in $q^+\,$.

\subsection{${\cal N}=2$ supersymmetric Yang-Mills theory}
The HSS formulation of ${\cal N}=2$ SYM theory reveals surprising affinities with the
ordinary (${\cal N}=0$) Yang-Mills theory.

The constraints \p{N2symC} defining the ${\cal N}=2$ SYM theory in the superspace
${\Bbb R}^{4\vert8}$ can be rewritten in HSS in the following equivalent
way
\be
\mbox{(a)} \;\{{\cal D}^{+}_\alpha, {\cal D}^{+}_\beta\} = \{\bar{\cal
D}^{+}_{\dot\alpha}, \bar{\cal D}^{+}_{\dot\beta}\} = \{{\cal
D}^{+}_{\alpha}, \bar{\cal D}^{+}_{\dot\beta}\} = 0\,, \quad \mbox{(b)} \;
[D^{++}, {\cal D}^{+}_\alpha] =[D^{++}, {\bar{\cal D}}^{+}_\alpha] = 0\,, \label{SYMconstr}
\ee
where ${\cal D}^{+}_\alpha = D^+_\alpha + ig A^+_\alpha\,$,
${\bar{\cal D}}^{+}_{\dot\alpha} = \bar D^+_{\dot\alpha} + ig \bar A^+_{\dot\alpha}\,$ and
$A^+_\alpha, \bar A^+_{\dot\alpha}$ are some spinor harmonic superfields with values in
the algebra of the gauge group, $g$ being the coupling constant. The equivalence of \p{SYMconstr}
to \p{N2symC} can be proven
quite analogously to the equivalence of \p{Constrq} and \p{HypC}. Eqs. (\ref{SYMconstr}b)
imply that ${\cal D}^{+}_\alpha = (D^l_\alpha + i A^l_\alpha)u^+_l\,$,
${\bar{\cal D}}^{+}_{\dot\alpha} = (\bar D^k_{\dot\alpha} + i \bar A^k_{\dot\alpha})u^+_k\,$,
then (\ref{SYMconstr}a) yield \p{N2symC} by taking off the harmonics $u^+_l u^+_k\,$.

The constraints (\ref{SYMconstr}a) are immediately recognized as integrability conditions for the existence of
the gauge-covariant version of Grassmann harmonic analytic superfields\footnote{A similar
interpretation of ${\cal N}=2$ SYM constraints as integrability conditions was given by A. Rosly \cite{Rosl}.}.
Once again, the advantage
of the new representation of the ${\cal N}=2$ SYM constraints can be
revealed by passing to the analytic basis in the superspace ${\Bbb H}^{4+2\vert8}$ and to a new
gauge frame, in which the covariant derivatives ${\cal D}^{+}_{\alpha}\,$,
$\bar{\cal D}^{+}_{\dot\beta}$ become ``short'', thus explicitly solving (\ref{SYMconstr}a):
\be
{\cal D}^{+}_{\alpha} = \partial^{+}_\alpha\,, \quad
\bar{\cal D}^{+}_{\dot\beta} = {\partial}^{+}_{\dot\beta}\,.
\ee
The existence of these basis and frame follows from the general Frobenius theorem.
At the same time, the harmonic derivative $D^{++}$ in the new frame acquires a non-trivial
gauge connection $V^{++}$,
\be
D^{++} \;\Rightarrow \; {\cal D}^{++} = D^{++} + ig V^{++}\,. \label{DcovD}
\ee
The constraint (\ref{SYMconstr}b) in the analytic  basis and frame is just the statement
that $V^{++}$ is the harmonic analytic ${\cal N}=2$ superfield
\be
V^{++} = V^{++}(\zeta, u)\,.
\ee

The harmonic analytic gauge connection $V^{++}$ is the fundamental gauge prepotential
of ${\cal N}=2$ SYM theory. It undergoes the following gauge transformations
\be
(V^{++})' = {1\over ig}
\mbox{e}^{i\omega}\left( D^{++} + ig V^{++} \right)\mbox{e}^{-i\omega}~,
\ee
where $\omega(\zeta,u)$ is an arbitrary analytic
gauge parameter. This transformation law resembles the standard gauge transformation
of the ${\cal N}=0$ Yang-Mills connection. The evident difference is that the latter covariantizes
the ordinary $x$-derivative, while $V^{++}$ covariantizes one of the two derivatives on the
internal harmonic sphere $S^2 \sim SU(2)_A/U(1)\,$.

The harmonic connection $V^{++}$ contains
infinitely many component fields in its combined $\theta, u$ expansion, like the
hypermultiplet superfield $q^+(\zeta, u)\,$. The difference from
$q^+(\zeta, u)\,$ lies, however, in the fact that almost all of these fields are
pure gauge degrees of freedom: they can be gauged away by $\omega (\zeta, u)$ which also
contains infinitely many components.
The finite reminder of $(8+8)$ components is just the off-shell
content of the superspin 0, superisospin 0 ${\cal N}=2$ vector multiplet.
More precisely, in the proper Wess-Zumino gauge $V^{++}$ has
the following form:
\bea V^{++}_{WZ} &=& (\theta^+)^2 w(x_A) + (\bar\theta^+)^2
\bar{w}(x_A) + i\theta^+\sigma^m\bar\theta^+ V_m(x_A)
+ (\bar\theta^+)^2\theta^{+\alpha} \psi_{\alpha}^i(x_A)u^-_i \nn \\
&+& (\theta^+)^2\bar\theta^+_{\dot\alpha}\bar\psi^{\dot\alpha i}(x_A)u^-_i +
(\theta^+)^2(\bar\theta^+)^2 D^{(ij)}(x_A)u^-_iu^-_j ~. \eea
Here, $V_m, w,\bar
w, \psi^{\alpha}_i, \bar\psi^{\dot\alpha i}, D^{(ij)}$ are the gauge field,
a complex physical scalar field, a  doublet of gaugini and a triplet of auxiliary
fields, respectively. All the geometric quantities of ${\cal N}=2$ SYM theory
(spinor and vector connections, covariant superfield strengths, etc), as well
as the invariant action, admit a concise representation in terms of
the fundamental geometric object $V^{++}(\zeta,u)$.  The details of how to construct the
invariant action can be found in the book \cite{Book} and in the original HSS papers.
In particular, the closed $V^{++}$ form of the ${\cal N}=2$ SYM action was proposed
for the first time in \cite{ZUP1}.

The $q^+$ hypermultiplet actions \p{ActqFr} or \p{333}  can be coupled to $V^{++}$ in a way quite
analogous to how one couples, e.g. fermions to the ordinary Yang-Mills field.
One should place the superfield $q^+$ into the appropriate representation of the gauge group
and assume for it the following transformation law
\begin{equation}
\delta
q^+_r(\zeta, u^\pm)= i\omega^A(\zeta, u^\pm)(T^A)_{rs}
q^+_s(\zeta, u^\pm)\;, \label{1.48}
\end{equation}
where $T^A$ are the generators of the gauge group in the given representation
and $\omega^A$ are the corresponding parameters. As usual, one should
covariantize the derivatives entering the action. In our case, this is the harmonic derivative in the $q^+$ action:
\begin{equation}
D^{++}\ \Longrightarrow\ {\cal D}^{++}=D^{++} +
iV^{++ A}T^A(\zeta,u)\;. \label{1.49}
\end{equation}

The general variety of theories in which ${\cal N}=2$ Yang-Mills fields interact with
hypermultiplets is known to contain a subclass of four-dimensional {\em ultraviolet
finite quantum field theories} (in particular, ${\cal N}=4$ Yang-Mills
theory). They also reveal remarkable properties of duality
\cite{aS200}. Harmonic superspace offers a unique possibility to
formulate these hybrid theories in a way that ${\cal N}=2$ supersymmetry stays manifest
off shell. For instance, the ${\cal N}=4$ super Yang-Mills theory action in the HSS
formulation is a sum of the action of $V^{++}$ and that
of $q^+$ in the adjoint representation, with the minimal coupling to $V^{++}$.
The HSS formulation considerably
simplifies many aspects and makes manifest many features of such
theories, e.g. the proof of non-renormalization theorems, finding
out the full structure of the quantum effective actions, etc.

The harmonic approach is also very appropriate for the description of
general non-minimal self-couplings of vector ${\cal N}=2$
supermultiplets. These theories are unique because they are the
only ${\cal N}=2$ supersymmetric field-theoretical models that admit a
natural chiral structure of interactions. For this reason they may
be useful in the phenomenological context as a possible basis of
${\cal N}=2$ GUT models. Sigma models inherent to these couplings are of
interest in their own right. Their target manifolds are of the
special K\"ahler type \cite{Gates,aC5} and have been discussed, e.g.,  in
connection with the so-called $c*$-map \cite{aC9}.

\setcounter{equation}{0}
\section{${\cal N}=2$ Einstein supergravity ``from scratch''}
The first example of off-shell ${\cal N}=2$ Einstein supergravity was derived at the component level
in \cite{VassFra, nider, nider2}. As was shown in \cite{aG16}, in the HSS formulation it corresponds to choosing the so called  nonlinear
multiplet as a compensator in conformal ${\cal N}=2$ supergravity. Actually, this kind of ${\cal N}=2$ Einstein SG can be
deduced in the HSS approach without referring to conformal SG \cite{GS1}, directly from the preservation of the harmonic Grassmann
analyticity and the appropriate choice of the fundamental gauge group, much like to ${\cal N}=2$ SYM
theory\footnote{An attempt in a similar direction was also undertaken in \cite{DelaKa}.}. Here we give a brief derivation of this theory
following ref. \cite{GS1}.

\subsection{From central to analytic bases}
The starting point of the consideration in \cite{GS1} is the so called $\tau$ gauge group acting on the ${\cal N}=2$ superspace
coordinates,
\bea
&& Z^{M} = \large(x^m, x^5, \theta^{\mu i}, \bar\theta^{\dot\mu k} \large)\,, \nn \\
&& \delta x^m = \tau^m(Z), \quad \delta x^5 =\tau^5(Z)\,, \quad \delta\theta^{\hat\mu i} = \tau^{\hat\mu i}(Z)\,,  \quad \hat\mu = (\mu, \dot\mu)\,.\lb{Tau}
\eea
The constraints of ${\cal N}=2$ Einstein SG can be then written as
\begin{eqnarray}
&& \{{\cal D}^i_\alpha\,, {\cal D}^k_\beta\} = \varepsilon_{\alpha\beta}\varepsilon^{ik} {\cal D}_5 + {\rm curvature} \quad ({\rm and \; c.c.}),\lb{SGconstr1} \\
&& \{{\cal D}^i_\alpha\,, \bar{\cal D}_{k\dot\beta}\} = \delta^{i}_{k} {\cal D}_{\alpha\dot\beta} + {\rm curvature}\,, \lb{SGconstr2}
\end{eqnarray}
where the indices $\alpha, \dot\beta, i$ of the spinor covariant derivatives are transformed by the local tangent space Lorentz group $SL(2, C)$ with
parameters $\Omega^{(\alpha\beta)}(Z), \bar\Omega^{(\dot\alpha\dot\beta)}(Z)$ and global
automorphism group $SU(2)_{aut}$ with parameters $\tau^{(ik)}$. The bosonic covariant derivative ${\cal D}_{\alpha\dot\beta}$ is the proper covariantization
of the derivative $\partial_m\,$. The gauge group parameters are assumed to be independent of $x^5$, so  ${\cal D}_5$
is constrained to reduce to the partial derivative
\be
{\cal D}_5 = \frac{\partial}{\partial x^5}\,. \lb{D5}
\ee
In fact, the actual constraints are only those parts of the relations \p{SGconstr1}, \p{SGconstr2},which are symmetric in the $SU(2)$ indices, {\it i.e.},
\be
\{{\cal D}^{(i}_{\hat\alpha}\,, {\cal D}^{k)}_{\hat\beta}\} = 0\,, \quad  \hat{\alpha} := (\alpha, \dot\alpha)\,. \lb{BasConstrCB}
\ee
The traces just yield the definition  of
${\cal D}_{\alpha\dot\beta}\,, {\cal D}_5$.

At the next step, harmonics come into play
\be
Z^{M} \rightarrow \{Z^M, u^{\pm i}\} = \{(x^m, x^5, \theta^{\hat\mu i}), u^{\pm i}\}\,.
\ee
One defines the harmonic projections of the spinor derivatives ${\cal D}^i_{\hat\alpha} \rightarrow {\cal D}^\pm_{\hat\alpha}$,
${\cal D}^\pm_{\hat\alpha} = u^\pm_i{\cal D}^i_{\hat\alpha}$ and
write the essential part of \p{SGconstr1}, \p{SGconstr2} as
\be
\{ {\cal D}^+_{\hat\alpha},  {\cal D}^+_{\hat\beta} \} = {\rm curvature}\,. \lb{SGconstr3}
\ee

Clearly, this is just the integrability condition for the existence of the covariantly analytic harmonic ${\cal N} = 2$ superfields
\be
{\cal D}^+_{\hat\alpha}\Phi (Z, u) = 0\,. \lb{AnalH}
\ee
Like in the case of ${\cal N} = 2$ SYM theory, this implies the existence of the basis where this analyticity gets manifest. To accomplish this,
we need to extend \p{SGconstr3}  to the so called ${\bf CR}$ (`Cauchy-Riemann'') structure \cite{CRStruc}, from which both the linearity of ${\cal D}^\pm_{\hat\alpha}$ in the harmonic
variables and the original constraints in the ``central basis'' follow. Such a structure is obtained by adding, to the relation \p{SGconstr3}, the following set
of the relations involving the harmonic derivatives
\bea
&& ({\rm a}) \;[{\cal D}^{++}, {\cal D}^+_{\hat\alpha}] = 0\,, \quad ({\rm b}) \; [{\cal D}^{0}, {\cal D}^+_{\hat\alpha}] = {\cal D}^+_{\hat\alpha}\,,\nn \\
&& ({\rm c})\; [{\cal D}^{0},{\cal D}^{++}] = 2 {\cal D}^{++}\,. \lb{HSSCR}
\eea
In the central basis, where ${\cal D}^{++} = D^{++} = u^{+i}\frac{\partial}{\partial u^{-i}}$ and ${\cal D}^{0} = D^0 = u^{+i}\frac{\partial}{\partial u^{+i}} -
u^{-i}\frac{\partial}{\partial u^{-i}}$, the constraints (\ref{HSSCR}a,b) indeed imply ${\cal D}^+_{\hat\alpha} = u^+_i{\cal D}^i_{\hat\alpha}$  and,
after substitution this in \p{SGconstr3} and taking off the harmonics, the original constraints \p{BasConstrCB} are recovered.

The main merit of singling out the above ${\bf CR}$ structure is the opportunity to pass to the new, analytic basis, in which just the spinor
derivatives ${\cal D}^+_{\hat\alpha}$ become short. This is achieved through introducing the proper ``bridges'' to the analytic superspace coordinates,
that allows one to reach  the ``almost simple'' form for ${\cal D}^+_{\hat\alpha}$
\be
{\cal D}^+_{\hat\alpha} = E^{\;\hat\mu}_{\hat\alpha}\frac{\partial}{\partial \theta_A^{-\hat\mu }} + A^+_{\hat\alpha} \equiv \nabla^+_{\hat\alpha} +  A^+_{\hat\alpha}\,,\lb{VielbE}
\ee
where $E^{\hat\mu}_{\hat\alpha}$ are the only remaining vielbeins. The analytic basis of the harmonic superspace in the curved case
is constituted by the coordinates
\be
\{Z_A, u^\pm_i\} = \{(x^m_A, x^5_A, \theta_A^{+\hat\mu },u^\pm_i), \theta_A^{-\hat\mu}\} \equiv (\zeta, x^5_A, u^\pm_i, \theta_A^{-\hat\mu}),
\ee
such that the curved analytic subspace, $(\zeta_, x_A^5, u^\pm_i) := (x^m_A, \theta_A^{+\hat\mu }, x^5_A, u^\pm_i)$, is preserved by the action of the analytic basis gauge group
\bea
&& \delta x^{m, 5}_A = \lambda^{m, 5}(z_A, u)\,, \;\; \delta \theta_A^{+\hat\mu } = \lambda^{+\hat\mu }(z_A, u)\,,\;\; \delta \theta_A^{-\hat\mu } =
\lambda^{\hat\mu -}(z_A, u, \theta_A^-)\,, \lb{analTran} \\
&& \delta u^\pm_i = 0\,. \lb{analu}
\eea
The indices $\hat\alpha$ of \p{VielbE} are still rotated by the local Lorentz parameters $\Omega^{(\alpha\beta)}(Z), \bar\Omega^{(\dot\alpha\dot\beta)}(Z)$ which are covariantly
$u^\pm_i$ independent,
$$
{\cal D}^{++}\Omega^{(\alpha\beta)} = {\cal D}^{++}\bar{\Omega}^{(\dot\alpha\dot\beta)}=0\,.
$$
So the vielbein $E^{\;\hat\mu}_{\hat\alpha}$ and the gauge connection $A^+_{\hat\alpha}$ in \p{VielbE} have the following transformation laws
\bea
&&\delta E^{\;\hat\mu}_{\hat\alpha} = \Omega_{\hat\alpha}^{\;\hat\beta}E^{\;\hat\mu}_{\hat\beta} + E^{\;\hat\nu}_{\hat\alpha}\partial^+_{\hat\nu} \lambda^{-\hat\mu}\,, \quad \partial^\pm_{\hat\nu} :=
\frac{\partial}{\partial \theta^{\mp\hat\mu}}\,, \lb{TranE} \\
&&\delta  A^+_{\hat\alpha\hat\beta\hat\gamma} = -\nabla^+_{\hat\alpha}\Omega^{\beta\gamma} + \Omega_{\hat\alpha}^{\;\hat\rho}A^+_{\hat\rho\hat\beta\hat\gamma}
+\Omega_{\hat\beta}^{\;\hat\rho}A^+_{\hat\alpha\hat\rho\hat\gamma} + \Omega_{\hat\gamma}^{\;\hat\rho}A^+_{\hat\alpha\hat\beta\hat\rho}\,. \lb{TranA}
\eea
Here,
\bea
\Omega_{\hat\alpha}^{\;\;\hat\beta} = \begin{pmatrix}
        \Omega^{\;\;\beta)}_{(\alpha} &\quad  0 \\
         0 &\quad  \bar\Omega^{\;\;\dot\beta)}_{(\dot\alpha}
    \end{pmatrix}.
\eea

In the analytic basis, the spinor derivative ${\cal D}^+_{\hat\alpha}$ becomes almost short, while the harmonic derivative ${\cal D}^{++}$ acquires non-trivial vielbeins,
\bea
&&{\cal D}^{++}_{AB} = \partial^{++} + H^{++m,5}\partial_{m, 5} + H^{++\hat\mu\pm}\partial_{\hat\mu}^{\mp} \equiv \partial^{++} + H^{++M}\partial_{M}\,,  \lb{AnalVielb}\\
&& \delta H^{++M} = {\cal D}^{++}\lambda^M\,, \quad M = (m, 5, \hat\mu{\pm})\,. \lb{TranAnalH}
\eea
The basic constraint (\ref{HSSCR}a) implies the following conditions for the harmonic vielbeins
\bea
&& \nabla^+_{\hat\alpha}H^{++m,5}\partial_{m, 5} + \nabla^+_{\hat\alpha} H^{++\hat\mu+}\partial_{\hat\mu}^{-} + ({\cal D}^{++}E^{\;\hat\mu}_{\hat\alpha} -
\nabla^+_{\hat\alpha} H^{++\hat\mu-})\partial_{\hat\mu}^{+} \nn \\
&& -\, {\cal D}^{++} A^{+}_{\hat\alpha} = 0. \lb{BasConstrH}
\eea
The obvious consequence is the analyticity of the basic vielbeins
\bea
&& \nabla^+_{\hat\alpha}H^{++m,5} = \nabla^+_{\hat\alpha} H^{++\hat\mu+} = 0\,\Rightarrow \; \nn \\
&&H^{++m,5} = H^{++m,5}(
\zeta, u)\,, \quad H^{++\hat\mu+} = H^{++\hat\mu+}(\zeta, u)\,.\lb{AnalitH}
\eea
As for $H^{++\hat\mu-}$, it can be gauged away into its flat limit, using the fact that the parameter $\lambda^{\hat\mu-}$ is a general harmonic superfunction,
\bea
H^{++\hat\mu-} = \theta^{+\hat\mu}_A\,, \quad {\cal D}^{++}\lambda^{-\hat\mu} = \lambda^{+\hat\mu}\,. \lb{Gauge1}
\eea

Now, recalling the gauge transformations \p{TranAnalH}, we can cast the remaining analytic vielbeins in the following Wess-Zumino form
\begin{eqnarray}
  &&H^{++m}(\zeta,u) = -2i\theta^+\sigma^a\bar\theta^+ e^m_a + \kappa(\bar\theta^+)^2
  \theta^{+\mu} \psi^m_{\mu i}u^{-i} \nn\\
  &&+\, \kappa(\theta^+)^2\bar\theta^+_{\dot\mu}
  \bar\psi_i^{m\dot\mu}u^{-i} + \kappa(\theta^+)^2(\bar\theta^+)^2 V^m_{ij}u^{-i}u^{-j}\;,\nn\\
&&H^{++\mu+}(\zeta,u) = \kappa(\theta^+)^2\bar\theta^+_{\dot\mu}(A^{\mu\dot\mu}+ iV^{\mu\dot\mu})
   + \kappa(\bar\theta^+)^2\theta^{+\nu} [\delta_\nu^\mu (M + i N) + T^{\;\mu)}_{(\nu}] \nn \\
&& +\,  \kappa(\theta^+)^2(\bar\theta^+)^2
   \xi^\mu_i u^{-i}\;, \quad H^{++\dot\mu+} = \widetilde{H^{++\mu+}} \nn\\
&&H^{++5}(\zeta,u) = i[(\theta^+)^2 - (\bar\theta^+)^2 ] + i\kappa \theta^+\sigma^a\bar\theta^+ B_a + \kappa (\bar\theta^+)^2 \theta^{+\mu}\rho^i_\mu u^-_i \nn \\
  && +\,\kappa (\theta^+)^2 \bar\theta^{+\dot\mu}\bar\rho^i_{\dot\mu} + \kappa (\theta^+)^2(\bar\theta^+)^2 S^{(ij)}u^-_i u^-_j\,,\label{WZMin}
\end{eqnarray}
$\kappa$ being Newton's constant ($[\kappa] = -1$). Here the fields $e^m_a, \psi^m_{\mu i}, \bar\psi^{m i}_{\dot\mu}, B_a$ describe the graviton, gravitini and ``graviphoton'' (gauge field
for the central charge local shifts), all in the ``frame-like'' formulation. All other fields are auxiliary.
With taking into account the residual gauge freedom of \p{WZMin} (local diffeomorphisms, local Lorentz transformations and local supersymmetry)
we are left with ${(40 + 40)}$ essential degrees of freedom, that is just the off-shell field content of the simplest ${\cal N}=2$ SG \cite{VassFra,nider}.

Our next point is singling out the remaining constraints from \p{BasConstrH}. Using \p{Gauge1},  the coefficient of $\partial^+_{\hat \mu}$ yields
\bea
{\cal D}^{++} E^{\;\hat\mu}_{\hat\alpha} = 0\,, \lb{ConstrE}
\eea
which means the covariant harmonic independence of $E^{\;\hat\mu}_{\hat\alpha}$. Because of this property one can accomplish the further gauge-fixing by making use
of the covariantly $u$-independent parameters $\Omega_{\alpha\beta}$.  The corresponding holomorphic and anti-holomorphic traceless  parts
of the diagonal in the matrix $E^{\;\hat\mu}_{\hat\alpha}$ can be gauged away , after which the vielbein can be reduced to the form
\begin{equation}\label{EFF}
E^{\;\hat\mu}_{\hat\alpha}
    =
    \begin{pmatrix}
        F\delta^{\;\mu}_\alpha &\quad F F^{\;\dot\mu}_\alpha \\
         \tilde{F} \tilde{F}^{\;\mu}_{\dot\alpha} &\quad  \tilde{F}\delta^{\;\dot\mu}_{\dot\alpha}
    \end{pmatrix}.
\end{equation}
Respectively, the Lorentz transformations in this gauge get induced by the coordinate ones
\be
\Omega_{\alpha\beta} =-\Delta^+_{(\alpha}\lambda^-_{\beta)}, \quad \Delta^+_{\alpha} := \frac1{F} \nabla^+_{\alpha}  = \partial^+_{\alpha} + F^{\;\dot\mu}_\alpha \bar\partial^+_{\dot\mu}\,.
\ee

Finally, \p{BasConstrH} yields the harmonic independence of the connection
\be
{\cal D}^{++}A^+_{\hat\alpha} = 0.
\ee
There still remains the constraint \p{SGconstr3}. The vanishing of the torsion in this anticommutator
amounts to the relations
\bea
({\rm a})\;{\cal D}^+_{(\alpha} E^{\hat\mu}_{\beta)} = 0\,, \quad ({\rm b})\;{\cal D}^+_{\alpha}E^{\hat\mu}_{\dot\beta}
+ \bar{\cal D}^+_{\dot\beta}E^{\hat\mu}_{\alpha} = 0  \quad ({\rm and \;c.c.})\,,  \lb{TorCon}
\eea
which allow one to express $A^+_{\hat\alpha}$ through
the entries of \p{EFF}
\bea
({\rm a})\;A^+_{\alpha\beta\gamma} =2 \varepsilon_{\alpha(\beta}\Delta^+_{\gamma)} F\,, \quad ({\rm b})\;A^+_{\alpha\dot\beta\dot\gamma} = - 2(\nabla^+_{\alpha} E_{(\dot\beta}^{\;\hat\mu})
E^{-1}_{\hat\mu\dot\gamma)} - 2 (\bar\nabla^+_{(\dot\beta} E_{\alpha}^{\;\hat\mu})E^{-1}_{\hat\mu\dot\gamma)}\,. \lb{Connect}
\eea
Indeed, substituting \p{EFF} into (\ref{TorCon}a) immediately yields (\ref{Connect}a) and also implies the constraint
\be
\Delta^+_{(\alpha} F_{\beta)}^{\;\mu} = 0 \; \Rightarrow \; \{\Delta^+_{\alpha}, \Delta^+_{\beta}\} = 0. \lb{ConstrF}
\ee

Note that after fixing the ``soldering'' gauge \p{EFF}, the actual difference between the tangent space and the ``world'' spinor indices becomes to some extent elusive,
so in what follows we will use the same letters to denote them.

\subsection{Covariant derivatives ${\cal D}^{--}$ and ${\cal D}^-_{\hat\alpha}$; further torsion constraints}
To finish the differential geometry routine, it remains to find the proper expressions through analytic potentials for
the quantities $F$ and $F_\alpha^{\;\ddot\mu}$ obeying the constraints \p{ConstrE} and \p{ConstrF}. This can be done after
defining the second (not preserving the analyticity) harmonic derivative, ${\cal D}^{--}$.

The basic relation defining ${\cal D}^{--}$ is the direct analog of the harmonic zero curvature condition of ${\cal N}=2$ SYM theory:
\bea
[{\cal D}^{++}, {\cal D}^{--}] = {\cal D}^0 = \partial^0
+ \theta^{+ \hat\mu}\partial^-_{\hat\mu} - \theta^{- \hat\mu}\partial^+_{\hat\mu}\,. \lb{BasD++D--}
\eea
After introducing the appropriate vielbeins in the analytic basis,
\bea
&&{\cal D}^{--}_{AB} = \partial^{--} + H^{--M}\partial_M =\partial^{--} + H^{--m,5}\partial_{m, 5} + H^{--\hat\mu\pm}\partial_{\hat\mu}^{\mp}\,, \nn \\
&& \delta H^{--M} =
{\cal D}^{--}\lambda^M\,, \lb{DefD--}
\eea
the relation \p{BasD++D--} gives rise to the equations
\bea
&& {\cal D}^{++}H^{--m,5} - {\cal D}^{--}H^{++m,5} = 0 \,, \nn \\
&&{\cal D}^{++}H^{--\hat\mu\pm} - {\cal D}^{--}H^{++\hat\mu\pm} = \pm \theta^{\pm\hat\mu}_A\,, \lb{SGZero}
\eea
which can be solved to express the negatively charged vielbeins in terms of the basic positively charged analytic ones.

The next important step is the definition of the second covariant spinorial derivative
\bea
{\cal D}^-_{\hat\alpha} = [{\cal D}^{--},  {\cal D}^+_{\hat\alpha} ]\,, \quad  [{\cal D}^{++},  {\cal D}^-_{\hat\alpha} ]  = {\cal D}^+_{\hat\alpha}\,.\lb{D-}
\eea
From eqs. \p{VielbE} and \p{DefD--} it then follows
\bea
{\cal D}^-_{\hat\alpha} = -\nabla^+_{\hat\alpha}H^{--\hat\mu\pm}\partial^{\mp}_{\hat\mu}
- \nabla^+_{\hat\alpha}H^{--m, 5}\partial_{m, 5} + A^-_{\hat\mu}\,, \quad  A^-_{\hat\mu} = {\cal D}^{--}A^+_{\hat\mu}\,,  \lb{StructD-}
\eea
where we used the property that eq. \p{ConstrE} implies ${\cal D}^{--}E^{\;\hat\mu}_{\hat\alpha} = 0$ (that can be easily proved in the central basis).
The vector connection is defined by the standard relation
\bea
{\cal D}_{\alpha\dot\alpha} = \frac12 \big( \{{\cal D}^+_\alpha, \bar{\cal D}^-_{\dot\alpha}\} - \{{\cal D}^-_\alpha, \bar{\cal D}^+_{\dot\alpha}\} \big).\lb{Vector}
\eea
It is convenient to rewrite the anticommutation relations involving ${\cal D}^-_{\hat\alpha}$ as
\bea
&& ({\rm a})\; \{{\cal D}_{\alpha}^-, {\cal D}_{\beta}^+\} = \varepsilon_{\alpha\beta}\partial_5 + {\rm curvature}\;\;({\rm and \;c.c.}); \nn \\
&& ({\rm b})\; \{{\cal D}_{\alpha}^{\pm} , \bar{\cal D}_{\dot\beta}^{\mp}\} = \pm{\cal D}_{\alpha\dot\beta} + {\rm curvature} \,.\lb{D+D-}
\eea

Now we are prepared to express the basic objects $F, F^{\dot\mu}_\alpha$ through the negatively charged vielbeins
$H^{--m, 5}$ and  $H^{--m\hat\mu \pm}$ and, further, through the fundamental analytic vielbeins by eqs. \p{SGZero}. For this purpose we introduce the matrices
\bea
&& \;e_{\hat\mu}^{\;\hat\nu} \;:= \partial^+_{\hat\mu} H^{--\hat\nu+}\,, \nn \\
&& e_{[\hat\mu\hat\nu]}^{\;\;m,5} :=
\partial^+_{\hat\mu}\partial^+_{\hat\nu}H^{-- m, 5}
- \partial^+_{\hat\mu}e_{\hat\nu}^{\;\hat\rho}e^{-1 \hat\lambda}_{\hat\rho}\partial^+_{\hat\lambda}H^{--m,5} \nn \\
&& \;\;\;\quad \;\; := \,\big(e^{m, 5}_{\mu\dot\nu}, \; e^{m, 5}\varepsilon_{\mu\nu}, \;
\bar{e}^{\,m, 5}\varepsilon_{\dot\mu\dot\nu}\big).\lb{DefMartr}
\eea

The relation (\ref{D+D-}b) does not produce an new torsion constraints, since it is conventional and just serves to define the covariant vector derivative ${\cal D}_{\alpha\dot\alpha}$.
The torsion constraints imposed by (\ref{D+D-}a) are as follows (the vanishing of the coefficients of $\partial_m, \partial_5$ and $\partial^-_\mu$, respectively)
\bea
&& E^{\hat\mu}_\alpha E^{\hat\nu}_\beta\, e^m_{[\hat\mu \hat\nu]} = 0\,, \lb{Constrm} \\
&& E^{\hat\mu}_\alpha E^{\hat\nu}_\beta\, e^5_{[\hat\mu \hat\nu]} = \varepsilon_{\alpha\beta}\,, \lb{Constr5} \\
&& E^{\hat\mu}_\alpha E^{\hat\nu}_\beta \, \partial^+_{\hat\mu} e_{\hat\nu}^{\;\hat\lambda} + {\cal D}^+_{[\alpha} E^{\;\hat\rho}_{\beta]}e_{\hat\rho}^{\;\hat\lambda} =0
\lb{ConstrAdd}
\eea
(and their conjugates). In \p{ConstrAdd} we used the constraint (\ref{TorCon}a). One more constraint following from (\ref{D+D-}a) (the vanishing of the coefficient of
$\partial^+_\mu$) can easily be shown to be valid as a consequence of \p{ConstrAdd}, while the latter itself proves also to be a consequence of the other ones.
Note that under the superdiffeomorphism group \p{analTran} different torsions are linearly rotated through each other, so only the full set of the equations
of the type \p{ConstrAdd} is covariant.  As an example,
let us give the transformation rule of  the left-hand side of eq. \p{ConstrAdd},
$${\cal F}^{+ \hat{\lambda}}_{\alpha\beta} := E^{\hat\mu}_\alpha E^{\hat\nu}_\beta \, \partial^+_{\hat\mu} e_{\hat\nu}^{\;\hat\lambda} + {\cal D}^+_{[\alpha} E^{\;\hat\rho}_{\beta]}e_{\hat\rho}^{\;\hat\lambda}.$$
We have
\bea
&&\delta {\cal F}^{+ \hat{\lambda}}_{\alpha\beta} = \Omega_\alpha^{\;\;\gamma}{\cal F}^{\hat{\lambda}}_{\gamma\beta} + \Omega_\beta^{\;\;\gamma}{\cal F}^{\hat{\lambda}}_{\alpha\gamma} +
 \Big[E^{\hat\mu}_\alpha E^{\hat\nu}_\beta\, e^m_{[\hat\mu \hat\nu]}\partial_m\lambda^{+ \hat\lambda} \nonumber \\
&& \quad\quad\quad\quad +\,{\cal F}^{+\hat{\gamma}}_{\alpha\beta}\,\big(\partial^-_{\hat\gamma}\lambda^{+\hat\lambda} + e^{-1\hat\rho}_{\hat\gamma}\partial^+_{\hat\rho} H^{--m}\partial_m\lambda^{+ \hat\lambda}\big)\Big]\,. \nonumber
\eea
We observe that ${\cal F}^{+\hat{\lambda}}_{\alpha\beta}$ homogeneously transform (as a tensor), provided that the constraint \p{Constrm} is valid.

The constraints serving to relate everything to the analytic potentials are just \p{Constrm} and \p{Constr5}.
In order to demonstrate this we define, following \cite{GS1},
\bea
&& f^{\mu\dot\nu} := e^m e_m^{\mu\dot\nu}\,, \quad  e^m_{\rho\dot\lambda} e_m^{\mu\dot\nu} = \delta^\mu_\rho \delta^{\dot\nu}_{\dot\lambda}\,,  \nn \\
&& f^5 := e^5 - e^m e_m^{\mu\dot\nu}e^5_{\mu\dot\nu}\,. \lb{Deff2}
\eea
Using \p{EFF}, we now find that the constraint \p{Constrm} amounts to the following equation
\bea
F^{\dot\nu}_\alpha = f^{\dot\nu}_\alpha + A \bar f^{\dot\nu}_\alpha\,, \quad A := \frac12 F^{\dot\nu\alpha} F_{\dot\nu\alpha}\,.\lb{UrFf}
\eea
Squaring this equation we obtain the quadratic equation for determining $A$:
\bea
&& A^2 \bar f^2 + 2 A(f\bar f - 1) + f^2 = 0\,, \lb{QuadrEq} \\
&& f^2 := f^{\dot\nu \alpha}f_{\dot\nu \alpha}\,, \; \bar f^2 := \bar f^{\dot\nu \alpha}\bar f_{\dot\nu \alpha} = \overline{(f^2)}\,,\;
f\bar f := f^{\dot\nu \alpha}\bar f_{\dot\nu \alpha}\,. \nonumber
\eea
A non-singular solution of \p{QuadrEq} is
\bea
A = f^2 \left[1 - f\bar f + \sqrt{(1 - f\bar f)^2 - f^2\bar f^2}\,\right]^{-1}\,. \lb{SolA}
\eea

Analogously, eq. \p{Constr5} is reduced to
\bea
F^{-2} = e^5 + A \bar{e}^5 - F^{\mu\dot\nu} e^5_{\mu\dot\nu}\,, \nn
\eea
whence, after using eqs. \p{UrFf} and \p{Deff2},
\bea
F = \frac1{\sqrt{f^5 + A\bar f^5}}\,. \lb{SolF}
\eea

\subsection{Building blocks and invariant action}
So far, we have succeeded to relate the components of \p{EFF} to the quantities $e^{m, 5}_{\mu\dot\nu},  e^{m, 5}, \bar{e}^{\,m, 5}$
which are expressed through the negatively charged vielbeins by eqs. \p{DefMartr} and, further, through the basic analytic vielbeins $H^{++ M}$, using
the harmonic flatness conditions \p{SGZero}. Note that the relations \p{UrFf}, \p{SolA} and \p{SolF} were derived in \cite{GS1} without directly
solving the torsion constraints \p{Constrm}, \p{Constr5}, based only upon the guessed transformation properties of the
involved quantities. In our presentation, all these transformation properties can be consistently deduced from the explicit form of
the matrix entries \p{DefMartr} and, further, from the definitions \p{Deff2}. The basic transformation laws from which all others can be derived are the following
\bea
&& \delta e^{\;\hat\rho}_{\hat\mu} = -\partial^+_{\hat \mu} \lambda^{-\hat\nu}e_{\hat\nu}^{\;\hat\rho} +
e^{\;\hat\omega}_{\hat\mu}\partial^-_{\hat \omega}\lambda^{+\hat\rho} + \partial^+_{\hat\mu} H^{--m}\partial_m \lambda^{+\hat\rho}\,, \nn \\
&& \delta e^{m,5}_{[\hat\mu\,\hat\nu]} = e^{n}_{[\hat\mu\,\hat\nu]} \,\lambda^{m, 5}_n - \partial^+_{\hat\mu}\lambda^{-\hat\omega}e^{m,5}_{[\hat\omega\,\hat\nu]},
+\partial^+_{\hat\nu}\lambda^{-\hat\omega}e^{m,5}_{[\hat\omega\,\hat\mu]}\,, \lb{TranBase}
\eea
where
\be
\lambda^{m, 5}_n  := \partial_n \lambda^{m, 5} - \partial_n\lambda^{+\hat\rho} e^{-1\,\hat\gamma}_{\hat\rho}\partial^+_{\hat\gamma} H^{-- m, 5}\,.
\ee

The basic use of the quantities constructed above are related to the possibility to construct the invariant volume of the curved harmonic superspace. In the analytic parametrization
it is defined as
\bea
&& \mu(Z) = du d^{12}Z\, E^{-1} \equiv  du d^4x_A d^4\theta^+_A d^4\theta^-_A\, E^{-1}\,, \quad E := {\rm Ber} E^{\;M}_A\,, \nn \\
&&\delta (du d^{12}Z) =  du d^{12}Z \large( \partial_m \lambda^m -\partial^+_{\hat\nu}\lambda^{-\hat\nu} - \partial^-_{\hat\nu}\lambda^{+\hat\nu}\large), \nn \\
&& \delta E = \large( \partial_m \lambda^m -
\partial^+_{\hat\nu}\lambda^{-\hat\nu} - \partial^-_{\hat\nu}\lambda^{+\hat\nu}\large) E\,,  \lb{TranBer}
\eea
where $E^{\;M}_A$ is a $12 \times 12$ matrix supervielbein entering the spinor and vector covariant derivatives \p{VielbE}, \p{StructD-} and \p{Vector}.
Instead of calculating the Berezinian $E$ directly, it is much simpler to construct this object from requiring it to have the transformation
property \p{TranBer}. The sought superdensity is given by the following expression
\be
E = \det (e^m_{\mu\dot\mu})\,\det{}^{-1} (e^{\;\hat\mu}_{\hat\nu})\;\sqrt{(1 - f\bar f)^2 - f^2\bar f^2}\,. \lb{Eexpr}
\ee
Using the transformation laws \p{TranBase} and their consequences, it is straightforward to check that the expression \p{Eexpr} indeed transforms
according to \p{TranBer}.

It was shown in \cite{GS1} that the invariant superfield action of the given version of ${\cal N}=2$ SG is given by the expression
\bea
S^{{\cal N}=2}_{SG} = -\frac{1}{\kappa^2} \int du d^4x_A d^4\theta^+_A d^4\theta^-_A\; E^{-1}\,H^{++ 5}H^{-- 5}\,. \lb{ActionE}
\eea
Its invariance under the local shifts of $x^m_A, \theta^{\pm\hat \mu}_A$ is evident, as $H^{++ 5}$, $H^{-- 5}$ are scalars with respect to these
transformations. To prove its invariance under gauge transformations with the analytic parameter $\lambda^5$ which do not act on the coordinates
in \p{ActionE} is a much more involved technical task. We invite the interested reader to consult the original paper \cite{GS1} and the book \cite{Book}.
Here we will limit our attention to the linearized version of this action following ref. \cite{Zup} and a recent paper \cite{BIZ1}.

\subsection{Linearized approximation}
The first step of the linearization consists in singling out the $\theta$ dependent backgrounds in the analytic potentials
\bea
&& H^{++ \alpha\dot\beta}  = -2i\theta^{+\alpha}\bar\theta^{+\dot\beta} + h^{++\alpha\dot\beta}\,, \nn\\
&& H^{++ 5} = i(\theta^{\hat{+}})^2 + h^{++ 5}\,, \quad (\theta^{\hat{+}})^2 := \big(\theta^{+\alpha}\theta^{+}_{\alpha}
- \bar\theta^+_{\dot\alpha}\bar\theta^{+ \dot\alpha}\big)\,, \nn \\
&& H^{++ \hat\mu +} =  h^{++ \hat\mu +}\,. \lb{Lin1Anz}
\eea
The negatively charged ``shifted'' potentials $h^{--\alpha\dot\beta, 5}$ are singled out from  $H^{-- \alpha\dot\beta, 5}$ by the analogous relations, with the background
pieces $-2i\theta^{-\alpha}\bar\theta^{-\dot\beta}$ and $i(\theta^{\hat{-}})^2$, respectively. Both sets of potentials are interrelated by the flatness conditions
following from the linearization of  eqs. \p{SGZero}:
\bea
&& D^{++}h^{--\alpha\dot\alpha} - D^{--}h^{++\alpha\dot\alpha}  + 2i\big( h^{--\alpha+}\bar\theta^{+\dot\alpha} + \theta^{+\alpha}h^{--\dot\alpha+}\big) = 0\,, \nn \\
&&D^{++}h^{--5} - D^{--}h^{++5} -2i\big( h^{--\alpha+}\theta^{+}_\alpha - \bar\theta^+_{\dot\alpha}h^{--\dot\alpha+} \big) = 0\,, \lb{m5} \\
&& D^{++}  h^{--\alpha+} - D^{--} h^{++\alpha+} =0\,, \quad D^{++}  h^{--\dot\alpha+} - D^{--} h^{++\dot\alpha+} =0\,, \nn \\
&& D^{++}  h^{--\alpha-} - h^{--\alpha+} =0\,, \quad D^{++}  h^{--\dot\alpha-} - h^{--\dot\alpha+} =0\,. \lb{spin}
\eea
These constraints are invariant under the following linearized form of the superfield gauge transformations defined in \p{TranAnalH} and \p{DefD--}
\bea
&&\delta_\lambda h^{\pm\pm \alpha\dot\alpha} = {D}^{\pm\pm } \lambda^{\alpha\dot\alpha} + 2i \big( \lambda^{\pm\alpha} \bar{\theta}^{\pm\dot{\alpha}}
        + \theta^{+\alpha}  \bar{\lambda}^{+\dot{\alpha}}\big)\,, \nn \\
&&\delta_\lambda h^{\pm\pm 5} = {D}^{\pm\pm} \lambda^5 - 2i \big(\lambda^{\pm{\alpha}} \theta^{\pm}_{\alpha} - \bar\theta^{\pm}_{\dot{\alpha}}\bar\lambda^{\pm\dot{\alpha}}\big),
\label{Linm5} \\
        && \delta_\lambda h^{++\hat{\alpha}+} = {D}^{++} \lambda^{+\hat{\alpha}}\,,  \lb{TranLin+} \\
&& \delta_\lambda h^{--{\hat\alpha}+} = {D}^{--} \lambda^{+{\hat\alpha}} - \lambda^{-\hat\alpha}\,,  \lb{TranLin-}\\
&& \delta_\lambda h^{--{\hat\alpha}-} = {D}^{--} \lambda^{-{\hat\alpha}}\,.  \lb{TranLin-3}
\eea
In all these formulas,
\bea
D^{\pm\pm} = \partial^{\pm\pm}
- 2i\theta^{\pm\alpha}\bar\theta^{\pm\dot\alpha}\partial_{\alpha\dot\beta} + \theta^{\pm \hat\mu}\frac{\partial}{\partial \theta^{\mp \hat\mu}}\,, \quad
\partial_{\alpha\dot\beta} := \sigma^m_{\alpha\dot\beta} \partial_m\,.
\eea

A peculiarity of the linearization limit considered is the unusual realization of the flat ${\cal N}=2$ supersymmetry on the potentials
$h^{\pm\pm m, 5}$\footnote{We denote by $\delta_\epsilon$
 so called passive transformations which differ from the more accustomed ``active'' transformations $\delta_\epsilon^*$
by the ``transport term'', $\delta_\epsilon^* = \delta_\epsilon - \delta_\epsilon Z^M \partial_M$.}
\bea
&& \delta_\epsilon h^{\pm\pm \alpha\dot\alpha} = -2i\big(h^{\pm\pm\alpha+}\bar\epsilon^{-\dot\alpha} + \epsilon^{-\alpha} h^{\pm\pm\dot\alpha +}\big)\,, \quad
\delta_\epsilon h^{\pm\pm 5} = 2i\,h^{\pm \hat\mu+}\epsilon^-_{\hat\mu}\,. \nn \\
&& \delta_\epsilon h^{++\hat\mu+} = \delta_\epsilon h^{--\hat\mu+} =\delta h^{--\hat\mu-}= 0\,. \lb{++susy}
\eea
It is straightforward to be convinced that eqs. \p{Linm5} -\p{TranLin-3} are covariant just under such modified transformation laws.
Actually, these transformation laws are valid in the full nonlinear case  too. The difference between the nonlinear and linearized cases is that in the former case
these rigid transformations form a subgroup of the gauge group \p{analTran} (and its counterpart for the negatively charged vielbeins),
while in the latter case they constitute an independent symmetry (which form a semi-direct product with the relevant linearized gauge transformations \p{TranLin-} and \p{TranLin-3}).

The next step in constructing the linearized invariant action is to define the important non-analytic superfields
\bea
&&G^{\pm\pm \alpha\dot\alpha} := h^{\pm\pm \alpha\dot\alpha} + 2i\big( h^{\pm\pm\alpha+}\bar\theta^{-\dot\alpha} + \theta^{-\alpha}h^{\pm\pm\dot\alpha+}\big)\,, \lb{Gpm} \\
&&G^{\pm\pm 5} := h^{\pm\pm 5} - 2i\,h^{\pm\pm\hat\alpha+}\theta^-_{\hat\alpha}\,. \lb{G5}
\eea
It is easy to check that the newly defined objects transform as the standard scalar ${\cal N}=2$ superfields
\be
\delta_\epsilon G^{\pm\pm \alpha\dot\alpha} = \delta_\epsilon G^{\pm\pm 5} = 0\,.
\ee
They also display simple transformation properties under the gauge transformations \p{TranLin+} - \p{TranLin-3}
\bea
&& \delta_\lambda G^{\pm\pm \alpha\dot\alpha} = D^{\pm\pm}\Lambda^{\alpha\dot\alpha}\,, \quad \delta_\lambda G^{\pm\pm 5} = D^{\pm\pm}\Lambda^5\,, \lb{TranG} \\
&& \Lambda^{\alpha\dot\alpha} = \lambda^{\alpha\dot\alpha} + 2i\big( \lambda^{+\alpha} \bar\theta^{-\dot\alpha} + \theta^{-\alpha}\bar\lambda^{+ \dot\alpha}\big)\,, \quad
\Lambda^5 = \lambda^5 - 2i \,\lambda^{+\hat\alpha}\theta^-_{\hat\alpha}, \lb{Lambda}
\eea
and satisfy the harmonic flatness conditions
\bea
D^{++}G^{-- \alpha\dot\alpha} = D^{--}G^{++ \alpha\dot\alpha}\,, \quad D^{++}G^{--5} = D^{--}G^{++5} \lb{FlatG}
\eea
as a direct consequence of the harmonic equations \p{m5} - \p{spin}. The invariant linearized action of ${\cal N}=2$ SG
can be constructed just from these objects.

Let us consider the manifestly ${\cal N}=2$ supersymmetric ``trial'' action
\be
S_1 = \int du d^4x d^8\theta \,G^{++\alpha\dot\alpha}G^{--}_{\alpha\dot\alpha}\,. \lb{S1}
\ee
Its gauge variation, with taking into account the relation \p{FlatG}, can be transformed to the expression
\be
\delta_{\lambda} S_1 = \frac12 \int du d^4xd^8\theta\, D^{--}\Lambda^{\alpha\dot\alpha}G^{++}_{\alpha\dot\alpha}\,.\lb{1step}
\ee
Now we pass to the integral over the analytic subspace,
\be
\int  dud^4xd^8\theta  = \int du d\zeta^{(-4)}  (D^+)^4\,, \quad (D^+)^4 = \frac{1}{16} (\bar D^+)^2 (D^+)^2\,,
\ee
and, after some algebra, using the property that both $\Lambda^{\alpha\dot\alpha}$ and $G^{++}_{\alpha\dot\alpha}$ are linear in $\theta^-_\alpha, \bar\theta^-_{\dot\beta}$
with analytic coefficients,
reduce this variation to
\be
\delta_{\lambda} S_1 = 2i \int  du d\zeta^{(-4)} \big( \partial_{\beta\dot\beta} \lambda^{+\beta}h^{++\dot\beta +} -
\partial_{\beta\dot\beta} \bar\lambda^{+\dot\beta}h^{++\beta +}\big). \lb{S11}
\ee

As the next step, we define
\be
S_2 = \int du d^4xd^8\theta  \,G^{++5}G^{--5}\,.
\ee
Applying similar manipulations, we find
\be
\delta_{\lambda} S_2= -2i \int  du d\zeta^{(-4)}  \big( \partial_{\beta\dot\beta} \lambda^{+\beta}h^{++\dot\beta +} -
\partial_{\beta\dot\beta} \bar\lambda^{+\dot\beta}h^{++\beta +}\big). \lb{S12}
\ee
So the sum
\be
S_{(s=2)} \sim -\big( S_1 + S_2 \big)  = -\frac{1}{\kappa^2}\int du d^4xd^8\theta \,\big(G^{++\alpha\dot\alpha}G^{--}_{\alpha\dot\alpha} + G^{++5}G^{--5} \big) \lb{Spin2N2}
\ee
is invariant under both rigid ${\cal N}=2$ supersymmetry and linearized gauge transformations (note the sign minus which is characteristic of the compensator actions). It is the invariant action of the linearized ${\cal N}=2$ SG and
the true ${\cal N}=2$ extension of the free spin 2 action. In the HSS approach it was firstly given (in a different form) in \cite{Zup}.

Note that the linearized ${\cal N}=2$ SG in the ordinary ${\cal N}=2$ superspace
was considered in \cite{Rivelles, Gates:1981qq} and, more recently, e.g., in \cite{Kuz2}. There, the approach based on the  Mezincescu-type prepotentials was applied and some other versions of ${\cal N}=2$ SG
were also considered. The present derivation uses the objects constructed  out of the fundamental analytic gauge potentials.

Finally, we briefly discuss how the correct component action for the spin 2 can be deduced from \p{Spin2N2}. The purely gravity parts of the potentials $G^{++ \alpha\dot\alpha}$ and $G^{++5}$,
in WZ gauge are given by the expressions
\begin{eqnarray}\label{G++ P}
&&G^{++\alpha\dot{\alpha}}(h) =
    -
    2i \kappa\theta^{+\beta} \bar{\theta}^{+\dot{\beta}} h_{\beta\dot{\beta}}^{\alpha\dot{\alpha}}
    +
    4 (\theta^+)^2 \bar{\theta}^{+\dot{\beta}} \bar{\theta}^{-\dot{\alpha}} B_{\dot{\beta}}^\alpha
    -
    4 (\bar{\theta}^+)^2 \theta^{+\beta} \theta^{-\alpha} {B}_\beta^{\dot{\alpha}}\,, \lb{ePart} \nonumber \\
&& G^{++5}(h) = -4 (\theta^+)^2  \bar{\theta}^{+\dot{\rho}} \theta^-_\mu B^\mu_{\dot{\rho}} - 4 (\bar{\theta}^+)^2 \theta^{+\mu} \bar{\theta}^-_{\dot{\rho}} \bar{B}_\mu^{\dot{\rho}}\,, \lb{Grav}
\end{eqnarray}
where
\begin{equation}\label{B field}
    B_{\mu\dot{\mu}}
    =
    \frac{\kappa}{4} \left(3 \partial_{\mu\dot{\mu}} h  - \partial^{\rho\dot{\rho}} h_{(\mu\rho)(\dot{\mu}\dot{\rho})}\right)\,, \quad
   h_{\alpha\beta\dot{\alpha}\dot{\beta}} = h_{(\alpha\beta)(\dot{\alpha}\dot{\beta})}
    +
    \varepsilon_{\alpha\beta} \varepsilon_{\dot{\alpha}\dot{\beta}} h
\end{equation}
and we gauged away the rest of components of $ h_{\alpha\beta\dot{\alpha}\dot{\beta}}$, {\it i.e.}, $\varepsilon_{\alpha\beta} h_{(\dot\alpha \dot\beta)},
\varepsilon_{\dot\alpha\dot\beta} h_{(\alpha \beta)}$ by the linearized local ``Lorentz'' transformations. After solving the relevant parts of the zero-curvature conditions
\p{FlatG} for $G^{--\alpha\dot{\alpha}}(h), G^{--5}(h),$ substituting all that in the action \p{Spin2N2} and doing there the $\theta$ integration, we obtain
\bea
S^{(s=2)}_{(h)} &=& -\int d^4x\; \Big[h^{(\alpha\beta)(\dot{\alpha}\dot{\beta})} \Box
        h_{(\alpha\beta)(\dot{\alpha}\dot{\beta})} -
        h^{(\alpha\beta)(\dot{\alpha}\dot{\beta})}
        \partial_{\alpha\dot{\alpha}} \partial^{\rho\dot{\rho}}
        h_{(\rho\beta)(\dot{\rho}\dot{\beta})} \nonumber \\
&&+\, 2\, h
        \partial^{\alpha\dot{\alpha}} \partial^{\beta\dot{\beta}}
        h_{(\alpha\beta)(\dot{\alpha}\dot{\beta})}
        - 6 h \Box h \Big].
\eea
It is the correct free action of the spin 2 fields $\big(h_{(\alpha\beta)(\dot{\alpha}\dot{\beta})}, \; h\big)$. It is invariant under the corresponding gauge transformations,
\be
\delta h_{(\alpha\beta)(\dot{\alpha}\dot{\beta})} =
        \frac{1}{\kappa}\, \partial_{(\alpha(\dot{\alpha}} a_{\beta)\dot{\beta})}\,,\qquad
        \delta h = \frac{1}{4 \kappa}\,\partial_{\alpha\dot{\alpha}} a^{\alpha\dot{\alpha}} \,,
\ee
where $a^{\alpha\dot{\alpha}}$ is a vector parameter of the linearized diffeomorphisms. For the spin 1 gauge field of the ${\cal N}=2$ spin
${\bf 2}$ multiplet the action \p{Spin2N2} also gives rise to the correct free Maxwell action\footnote{The correct sign in this case is restored after elimination of the symmetric tensor
auxiliary fields $T^{(\alpha\beta)}, \bar{T}^{(\dot\alpha\dot\beta)}$ which are present in $G^{\pm\pm \alpha\dot\alpha}, G^{\pm\pm 5}$ and have the same dimension as the Maxwell
field strength \cite{BIZ1}.}.

\setcounter{equation}{0}
\section{Conformal ${\cal N}=2$ supergravity}
Like in the ${\cal N}=1$ case, the basic object of various versions of Einstein ${\cal N}=2$ supergravity
is the irreducible multiplet of conformal ${\cal N}=2$ supergravity - the superspin 1, superisospin 0
${\cal N}=2$ Weyl multiplet. Any version of Einstein ${\cal N}=2$ supergravity can be obtained
as a theory of the appropriate matter compensator in the background of this ${\cal N}=2$ Weyl
multiplet \cite{Struct}.

\subsection{${\cal N}=2$ Weyl multiplet in HSS}
In a nice analogy with ${\cal N}=1$ conformal supergravity (see Sect. 3.2),
in the HSS approach the ${\cal N}=2$ Weyl multiplet is closely related to the fundamental
group of conformal ${\cal N}=2$ supergravity and has a clear geometrical meaning. The
underlying group is the group of general diffeomorphisms of the harmonic analytic superspace
$\left(\zeta^M, u^{\pm i}\right)$, while the Weyl multiplet is accommodated by
the harmonic analytic vielbeins which covariantize the analyticity-preserving harmonic
derivative $D^{++}$ with respect to this group. So the group-theoretical basis of ${\cal N}=2$
supergravity is the preservation of ${\cal N}=2$ harmonic analyticity in the curved case,
much like the preservation of ${\cal N}=1$ chirality provides the basis of ${\cal N}=1$
supergravity. An essential difference from the group of Einstein SG \p{analTran}, \p{analu}  is the presence of local $SU(2)$
symmetry and the corresponding non-trivial transformations of the harmonic variables:
\begin{eqnarray}
\delta x^m &=& \lambda^m(\zeta,u)\;,\nn \\
\delta \theta^{+\hat\mu}
&=& \lambda^{+\hat\mu}(\zeta,u)\;, \nonumber \\
\delta \theta^{-\hat\mu} &=& \lambda^{-\hat\mu}(\zeta, \theta^{-\hat\mu}, u), \nn \\
\delta u^+_i
&=& \lambda^{++}(\zeta,u) u^-_i\;, \quad  \delta u^-_i = 0\;,
\label{analDiff}
\end{eqnarray}
where as before $\hat\mu := (\mu, \dot\mu)$. The local parameters $\lambda$ in \p{analDiff}, except for $\lambda^{-\hat\mu}(\zeta, \theta^{-\hat\mu}, u)$  are arbitrary
analytic harmonic functions. Note that only the harmonics $u^+$
but not $u^-$ transform, yet preserving the harmonic defining relation
$u^{+i}u^-_i =1 \,$. This peculiarity is related to the special
realization of the rigid ${\cal N}=2$ superconformal group $SU(2,2|2)$ in the analytic superspace
(see \cite{Book}). The transformations \p{analDiff} provide gauging of this rigid group.
Using the standard Lie-bracket formalism, it is straightforward to make sure that the transformations \p{analDiff} indeed form a group.
In particular,
\bea
[\delta_1,\delta_2]u^+_i = \{(\lambda^m_1\partial_m+\lambda^{+\hat\mu}_1\partial^-_{\hat\mu} +\lambda^{++}_1\partial^{--})\lambda^{++}_2\}u^-_i -
(1\leftrightarrow 2).
\label{8.13.1}
\eea
We see that the commutator produces a transformation of $u^+_i$ of the same
type as in \p{analDiff}, with a new analytic parameter.

The starting point of the actual construction of conformal ${\cal N}=2$ supergravity is the
conformally invariant free action of the hypermultiplet in the flat HSS \cite{Book}:
\be
S= \frac12 \int du d\zeta^{(-4)}\; q^{+a}D^{++}q^+_a, \ \ \ a=1,2.
\label{8.12.1a}
\ee
Note the ``wrong'' sign of this action. The rigid conformal supergroup
$SU(2,2\vert 2)$ leaves the analytic superspace $(\zeta_A,u)$ invariant (see
\cite{Book}), while $D^{++}$ has an unusual transformation law
\bea
\delta D^{++} = -\lambda^{++}_{rig} D^0\,, \lb{Rig}
\eea
where $\lambda^{++}_{rig}$ is the ``flat superspace'' limit of the gauge super parameter $\lambda^{++}$ in  \p{analDiff}.
The lagrangian in \p{8.12.1a} is invariant under \p{Rig} just because of the algebraic property $q^{+ a}q^+_a = 0$. The integration measure
in \p{8.12.1a} is not invariant, but its transformation can be compensated by the appropriate coordinate-dependent rescaling of $q^{+ a}$,
once again exploiting the identical vanishing of $q^{+ a}q^+_a$. The details related to the analytic superspace preserving realization of
the supergroup $SU(2,2\vert 2)$ can be found in \cite{Book}.

The covariantization of the action \p{8.12.1a} amounts to covariantization of the flat harmonic derivative
\bea
&& D^{++} \rightarrow {\cal D}^{++}\,, \nn \\
&& {\cal D}^{++} = \partial^{++} +
H^{+4}\partial^{--}
+ H^{++\,m}\partial_m + H^{++\,\hat\mu +}
\partial^-_{\hat\mu} + H^{++\,\hat\mu -}\partial^+_{\hat\mu}\,,\lb{CovD++}
\eea
where all the coefficients of the supervielbein, except for $H^{++\,\hat\mu -}$, are analytic superfields, $H^{++M} = H^{++M}(\zeta, u), H^{+ 4} = H^{+4}(\zeta, u)$, $M := (m, \,\hat{\mu}+)$.
This analyticity is necessary in order to preserve the analyticity of the Lagrangian in \p{8.12.1a}.  Though $H^{++\,\hat\mu -}$ drops out from the covariantized $q^{+}$
action because of analyticity  of the superfield $q^+$, it should in general be  included, having in mind that  ${\cal D}^{++}$ should be as well covariant when
acting on non-analytic harmonic superfields. The natural generalization of the flat transformation law \p{Rig},
\bea
\delta {\cal D}^{++} = -\lambda^{++} D^0\lb{Loc}
\eea
implies the following transformation properties for the supevielbeins
\bea
&& \delta H^{++\,M} = {\cal D}^{++}\lambda^M -
\delta^{M}_{+\hat\mu}\theta^{+\hat\mu}\lambda^{++}~, \quad \delta H^{++\,\hat\mu -} = {\cal D}^{++}\lambda^{-\hat\mu}  +
\theta^{-\hat\mu}\lambda^{++}\,, \nn \\
&& \delta H^{+4}
= {\cal D}^{++}\lambda^{++}~,  \label{Vielbein}\\
&&D^0 \equiv \partial^0
+ \theta^{+\hat\mu} {\partial^{-}_{\hat\mu}} - \theta^{-\hat\mu} {\partial^{+}_{\hat\mu}}~. \nonumber
\eea
The non-analytic gauge superparameter $\lambda^{-\hat\mu}(\zeta, \theta^{-\hat\mu}, u)$ contains enough component parameters in order
to gauge the non-analytic vielbein coefficient $H^{++\,\hat\mu -}$ into its flat limit,
\bea
H^{++\,\hat\mu -} = \theta^{+\hat\mu}\,, \quad \Rightarrow \quad {\cal D}^{++}\lambda^{-\hat\mu} = \lambda^{+\hat\mu} - \theta^{-\hat\mu}\lambda^{++}\,. \label{GaugeConf}
\eea

The covariantized hypermultiplet action,
\bea
S_{cov}=\int du\,  d\zeta^{(-4)}\; q^{+a}{\cal D}^{++}q^+_a, \ \ \ a=1,2,
\label{8.12.1}
\eea
is invariant under \p{analDiff}, \p{Loc} by the same token as in the flat case under superconformal group, just due to the property $q^{+ a}q^+_a = 0$.
Indeed, the transformation \p{Loc} leaves the action invariant because of the evident property $D^0 q^{+ a} = q^{+ a}$, while the transformation of the analytic integration measure,
\bea
&&\delta (du d\zeta^{(-4)}) = (du d\zeta^{(-4)})\,\Lambda\,, \nn \\
&&\Lambda :=  {\partial_m \lambda^m } + \partial^{--}\lambda^{++}
- {\partial^{-}_{\hat\mu}}\lambda^{+\hat\mu}\,, \lb{TranMeas}
\eea
can be canceled by the appropriate analytic rescaling of $q^{+ a}$,
\bea
\delta q^{+ a} = -\frac12 \Lambda q^{+ a}\,. \lb{Tranlocq}
\eea

The vielbein coefficients $H^{++M}, H^{++++}$ are unconstrained analytic superfields involving infinite numbers of the component
fields which come from the harmonic expansions. Most of these fields,
like in the ${\cal N}=2$ SYM prepotential $V^{++}$, can be gauged away by the
analytic parameters $\lambda^M, \lambda^{++}$, leaving in the Wess-Zumino gauge just the
irreducible off-shell $(24+24)$-component ${\cal N}=2$ Weyl multiplet \cite{Weyl,nider2,nider3}:
\begin{eqnarray}
  &&H^{++m}(\zeta,u) = -2i\theta^+\sigma^a\bar\theta^+ e^m_a(x) + (\bar\theta^+)^2
  \theta^{+\mu} \psi^m_{\mu i}(x)u^{-i} \nn\\
  &&\phantom{H^{++m}(\zeta,u)=} + (\theta^+)^2\bar\theta^+_{\dot\mu}
  \bar\psi_i^{m\dot\mu}(x)u^{-i} + (\theta^+)^2(\bar\theta^+)^2 V^m_{ij}u^{-i}u^{-j}\;,\nn\\
  &&H^{++\mu+}(\zeta,u) = (\theta^+)^2\bar\theta^+_{\dot\mu}A^{\mu\dot\mu}(x)
   + (\bar\theta^+)^2\theta^{+}_\nu t^{(\nu\mu)}(x) +  (\theta^+)^2(\bar\theta^+)^2
   \chi^\mu_i(x)u^{-i}\;,\nn\\
  &&H^{++\dot\mu+}(\zeta_A,u) = \widetilde{H^{++\mu+}} \;,\nn\\
  &&H^{+4}(\zeta,u) = (\theta^+)^2(\bar\theta^+)^2 D(x) \;. \label{8.17.2}
\end{eqnarray}
Here $e^m_a, \psi^m_{\mu i}, \bar\psi^{m i}_{\dot\mu},  V^{m}_{ij}, A^{\mu\dot\mu}$ are the
conformal graviton, gravitini and gauge fields for the local $SU(2)$ and
$\gamma_5$ transformations; all other fields are auxiliary.

\subsection{Covariant derivative ${\cal D}^{--}$ and minimal superconformal action}
Now, armed with the superfield differential geometry formalism for Einstein ${\cal N}=2$ SG described in Section 5, we can complete the geometric
setup of conformal ${\cal N}=2$ SG.

First, we define the second harmonic derivative ${\cal D}^{--}$
\bea
{\cal D}^{--} = \partial^{--}_A + H^{--m}\partial_m^A + H^{--\hat\mu\pm}\partial^{\mp A}_{\hat\mu}\,. \lb{D--Conf}
\eea
To deduce the transformation rule of ${\cal D}^{--}$, we first note that under $\lambda^{++}$ transformation, with the supervielbein $H^{--M}$ in \p{D--Conf} still untouched,
${\cal D}^{--}$ transforms as
\bea
\hat{\delta}_{\lambda^{++}}{\cal D}^{--} = -({\cal D}^{--}\lambda^{++})\partial^{--}_A \,.
\eea
Then the full covariant transformation law of ${\cal D}^{--}$ (with the proper transformations of the supervielbein coefficients) can naturally be chosen as
\bea
{\delta}{\cal D}^{--} = -({\cal D}^{--}\lambda^{++}) {\cal D}^{--}\,. \lb{D--Conf}
\eea
The appropriate transformation rules of $H^{--M}$ are then as follows
\bea
\delta H^{--M} = -({\cal D}^{--}\lambda^{++})H^{--M} + {\cal D}^{--}\lambda^M\,. \lb{H--Tran}
\eea
To ensure covariance of  the harmonic flatness condition, we are led to modify it as
\bea
[{\cal D}^{++} - H^{+4}{\cal D}^{--}, {\cal D}^{--}] = D^0\,.
\eea

In principle, one can now repeat the same steps as in the case of Einstein ${\cal N}=2$ SG: define the covariant spinor and vector derivatives, super torsions and
curvatures,  etc. This formalism could be of use, e.g., for solving the problem of constructing higher-derivative superconformal invariants and  related problems.
Here we will be interested only in constructing some off-shell actions of compensators in the background of ${\cal N}=2$ Weyl multiplet. Using them, one can derive the actions of
diverse versions of Einstein ${\cal N}=2$ SG including the one we have discussed in the previous Section.

It turns out that it is easy to generalize the action \p{ActionE} to the superconformal case. The relevant density $E$ defining the invariant integration measure
is still given by the same formula \p{Eexpr} but the objects entering it have different transformation properties as compared to those inherent to Einstein SG case,
eqs. \p{TranBase}, because of the presence of extra analytic gauge parameter $\lambda^{++}$. It is straightforward to find the new terms in the transformations of
$e^m_{[\hat\mu\,\hat\nu]}$ and
$e^{\;\hat\nu}_{\hat\mu}$
\bea
\delta_{\lambda^{++}}e^m_{[\hat\mu\,\hat\nu]} &=& -({\cal D}^{--} \lambda^{++}) e^m_{[\hat\mu\,\hat\nu]} \nn \\
&& -\, e^n_{[\hat\mu\,\hat\nu]}\big[ \partial_n\lambda^{++} \big(H^{--m} -
H^{--\hat\mu+}e^{-1 \hat\nu}_{\hat\mu} \partial_{\hat\nu}^+H^{--m}\big)\big], \lb{Dop1} \\
\delta_{\lambda^{++}}e^{\;\hat\nu}_{\hat\mu} &=& -({\cal D}^{++} \lambda^{++})e^{\;\hat\nu}_{\hat\mu}\nn \\
&&-\, \big(\partial_{\hat\mu}^+H^{--m}\partial_m\lambda^{++}
+ e^{\;\hat\rho}_{\hat\mu}\partial_{\hat\rho}^-\lambda^{++}\big)H^{--\hat\nu+}. \lb{Dop2}
\eea
The second piece in the transformations \p{H--Tran} has the same form as in Einstein ${\cal N}=2$ SG, so it produces the same weight transformation of $E$ as
in eq. \p{TranBer}. So we are led to consider  the impact of the transformations \p{Dop1}, \p{Dop2} only on various determinant factors in \p{Eexpr}. It is easy to check that
the object $f^{\mu\dot\mu} = e^m e_m^{\mu\dot\mu}$ remains invariant under the second, rotational part of the transformation \p{Dop1}, so the square root in
the definition \p{Eexpr} is also invariant. Then it is easy to show that
\bea
\delta_{\lambda^{++}} E = \big(\partial_A^{--} \lambda^{++} -{\cal D}^{--} \lambda^{++} \big) E\,.
\eea
Hence, the total transformation of $E$ reads
\bea
\delta_{\lambda} E = \big(\partial_m \lambda^m  -\partial^+_{\hat\nu}\lambda^{-\hat\nu} - \partial^-_{\hat\nu}\lambda^{+\hat\nu} + \partial_A^{--} \lambda^{++} -{\cal D}^{--} \lambda^{++} \big) E\,.
\eea
One should also take into account that the transformation of the integration element $dud^{12}Z$ gets now a contribution from the variation of the harmonics $u^+_i$, namely,
$\frac{\partial \lambda^{++} u^-_i}{\partial u^+_i} = \partial^{--}_A \lambda^{++}$, so that
\bea
\delta ( dud^{12}Z) =  du d^{12}Z \big( \partial_m \lambda^m  + \partial^{--}_A \lambda^{++}
-\partial^+_{\hat\nu}\lambda^{-\hat\nu} - \partial^-_{\hat\nu}\lambda^{+\hat\nu}\big). \lb{TranConfMeas}
\eea
Finally, we obtain
\bea
\delta (du d^{12}Z\,E^{-1}) = (du d^{12}Z \,E^{-1})\,({\cal D}^{--} \lambda^{++})\,.
\eea

To establish a contact with he previous Section, we need to extend the set of the harmonic superspace coordinates by $x^5$ and, correspondingly, extend the harmonic covariant
derivatives ${\cal D}^{\pm\pm}$ by the extra vielbeins $H^{\pm\pm 5}$,
\bea
{\cal D}^{++} \,\Rightarrow \, {\cal D}^{++} + H^{++ 5}\partial_5\,, \quad {\cal D}^{--} \,\Rightarrow \, {\cal D}^{--} + H^{-- 5}\partial_5\,.  \lb{ExtendDconf}
\eea
As before, nothing depends on $x^5$, $H^{++ 5}$ is analytic,  $H^{++ 5} = H^{++ 5}(\zeta, u)$, while $H^{-- 5}$ is related to $H^{++ 5}$ by the proper harmonic flatness equation.
In accord with the transformation properties \p{Loc}, \p{D--Conf}, $H^{\pm\pm 5}$ transform as
\bea
\delta H^{++ 5} = {\cal D}^{++}\lambda^5\,, \quad \delta H^{-- 5}  = -({\cal D}^{--}\lambda^{++})H^{--5} + {\cal D}^{--}\lambda^5\,.
\eea
Thus we observe that the same action
\bea
S^{{\cal N}=2}_{SG} = -\frac{1}{\kappa^2} \int du d^4x_A d^4\theta^+_A d^4\theta^-_A\; E^{-1}\,H^{++ 5}H^{-- 5} \lb{ActionA}
\eea
is invariant under the full superconformal group \p{analDiff} and $\lambda_5$ transformations,
\bea
\delta_\lambda S^{{\cal N}=2}_{SG} = 0\,, \quad \delta_{\lambda^5} S^{{\cal N}=2}_{SG} = 0\,.
\eea

\subsection{From conformal to Einstein ${\cal N}=2$ SG}
At this stage we have what is called ``minimal off-shell representation'' \cite{MinRepr,Struct}.  It involves $(32 + 32)$ off-shell degrees of freedom: $(24 + 24)$ from Weyl
${\cal N}=2$ multiplet and $(8 + 8)$ from the Maxwell multiplet described by $H^{++5}$. The scalar fields in $H^{++ 5}$ can serve as compensators for the scale and $R$-symmetry present
in superconformal  gauge group. However, the ${\cal N}=2$ superconformal action \p{ActionA}, though being formally identical
to that of ${\cal N}=2$ Einstein SG , eq. \p{ActionE}, involves a wider set of fields and a wider set of invariances. As it stands, it is in fact inconsistent and therefore cannot serve
as the appropriate action neither for ${\cal N}=2$ Einstein SG nor for ${\cal N}=2$ superconformal SG. The reason was
explained in detail in \cite{Book}. In brief, the component action in WZ gauges for Weyl and Maxwell supermultiplets contains an unwanted term
\be
\sim \int d^4 x\, D\varphi^2, \lb{Trouble}
\ee
where $D$ is the auxiliary field from $H^{+4}$ (see eq. \p{8.17.2}) and $\varphi$ is the imaginary part of the complex scalar field of $H^{++ 5}$.  Then varying
$D$ in \p{Trouble} yields meaningless constraint $\varphi^2 = 0$ \footnote{The non-existence of consistent invariant action  for the ``minimal representation''
in the component approach was shown in  \cite{MinRepr}.}. To evade this trouble, one needs one more compensating superfield besides $H^{++5}$, so as to
compensate the local $SU(2)$ with the analytic parameter $\lambda^{++}(\zeta, u)$.

The simplest variant is to consider an extended system involving the so-called nonlinear supermultiplet \cite{Struct}. In the flat HSS it is described by the analytic superfield $N^{++}$ obeying
the constraint \cite{Book}
\bea
D^{++}N^{++} + (N^{++})^2 = 0 \,.\lb{NonlMflat}
\eea
It contains $8 + 8$ independent off-shell components. The curved covariant generalization of \p{NonlMflat} is as follows
\bea
{\cal D}^{++}N^{++} + (N^{++})^2 - H^{+4} = 0 \,,\lb{NonlMflat1}
\eea
provided that $N^{++}$ has the following transformation law
\bea
\delta N^{++} = \lambda^{++}\,. \lb{TranN}
\eea
 One observes that the covariantized constraint \p{NonlMflat1} can serve as the definition of $H^{+4}$
\bea
H^{+4} = {\cal D}^{++}N^{++} + (N^{++})^2\,.
\eea
Furthermore, the transformation law \p{TranN} tells us that one can make use of the super parameter
$\lambda^{++}(\zeta, u)$ to entirely gauge away $N^{++}$
\bea
N^{++} = 0\; \Rightarrow \;H^{+4} = 0\,. \lb{NHgauge}
\eea
In this gauge the  superconformal diffeomorphisms \p{analDiff} are reduced to \p{analu}, the covariant derivatives ${\cal D}^{\pm\pm}$ to those defined
in the previous Section and the action \p{ActionA} to the previously given ${\cal N}=2$ SG action \p{ActionE}. Of course, the possibility to choose the gauge \p{NHgauge} in
\p{ActionE} is ensured by the superconformal invariance of this action. The trouble mentioned earlier is automatically resolved, as $D=0$ in the gauge \p{NHgauge}. Note that
no superconformally invariant action of $N^{++}$ alone can be constructed  even in the flat case. The same is true also in the curved case, and so the version of ${\cal N}=2$ Einstein SG
with $N^{++}$ as a compensator is described by the superconformal Maxwell action \p{ActionA}, with the additional  superconformally covariant constraint \p{NonlMflat1} imposed ``by hand''.

It is possible to choose as second compensator other ${\cal N}=2$ matter supermultiplets with finite numbers of the off-shell components, for which superconformally invariant actions exist in the flat HSS.
Such are the so called ``improved'' tensor ${\cal N}=2$ multiplet with the off-shell content $8 + 8$ \cite{Tensor} and the hypermultiplet with the non-trivially realized operator central charge \cite{CentrCharg1}. The HSS
formulation of these off-shell multiplets were given in \cite{Book} and \cite{CentrCharg3, CentrCharg2}.   All these multiplets, like $N^{++}$,  are described by analytic
superfields subjected to some covariant differential constraints. The superfield actions of the relevant versions of Einstein  ${\cal N}=2$ SG are sums of the superconformal
Maxwell action \p{ActionA} and the actions of the compensating matter superfields in the background of ${\cal N}=2$ Weyl multiplet. The aforementioned trouble with the auxiliary
field $D$ is solved in all cases due to the property that the field $D$ serves as a Lagrange
multiplier identifying some scalar fields from the matter compensator with the scalar field $\varphi$ from the Maxwell compensator action. In the appropriate gauge the remaining scalar
field $\phi$ plays the role of dilaton and the relevant part of the gravity sector of the full action takes the form
\be
S_{grav} = \int d^4 x \sqrt{-g}\Big[3\phi\Big( \Box - \frac16 R\Big)\phi + \xi^2 \phi^4\Big],
\ee
where $\Box = \nabla^m\partial_m$ and the second term corresponds to the cosmological constant. This action is invariant under local rescalings
\be
\delta g_{mn}(x) = 2a(x) g_{mn}(x), \delta \phi(x)  = -a(x) \phi(x)\,.
\ee
Fixing this extra gauge invariance as $\phi(x) = \kappa^{-1}$, we reproduce the standard Einstein action
\be
S_{Ein} = \int d^4 x \sqrt{-g} \Big( - \frac{1}{2\kappa^2} R + \frac{\xi^2}{\kappa^4}\Big).
\ee
This mechanism \cite{FrTs} is the ${\cal N}=0$ prototype of the general compensating procedure and it works in all known off-shell versions of ${\cal N}=2$ supergravity.

The  versions of ${\cal N}=2$ Einstein SG with the $8 + 8$ compensators mentioned above  exhibit the sets of $40 + 40$ off-shell components in WZ gauge, like in the minimal version with nonlinear multiplet $N^{++}$
as a compensator. The locally superconformal
action of the improved tensor multiplet in HSS \cite{GIOshw} is given by
\bea
S_{impr} \sim \int du\, d\zeta^{(-4)}\Big[(g^{++})^2 - \Gamma^{++} g^{++} c^{+-} - H^{+4}(1 + 2 g^{++} c^{--})\Big], \label{Impr}
\eea
where
\bea
g^{++} = \frac{{\ell}^{++}}{1 + \sqrt{1 + \ell^{++} c^{--}}}, \quad \ell^{++} = L^{++} - c^{++}, \;c^{\pm\pm} = c^{ik}u^\pm_i u^\pm_k
\eea
and the following covariant off-shell constraint and the transformation properties hold
\bea
&& \Big({\cal D}^{++} + \Gamma^{++} \Big) L^{++} = 0\,, \quad \Gamma^{++} := (-1)^{P(M)}\partial_M H^{++ M}\,,   \label{Gamma} \\
&& \delta\Gamma^{++} = 2 \lambda^{++} + {\cal D}^{++}\Lambda\, , \quad \delta L^{++} = -\Lambda\,L^{++}\,, \label{TranL}
\eea
where $\Lambda$ was defined in \p{TranMeas}. The superfield $L^{++}$ involves $8 + 8$ off-shell components, which, together with $32 + 32$ components of the ``minimal off-shell representation'',
yield total of $40 + 40$ off-shell components.

Analogously, the compensating hypermultiplet of ref.\cite{CentrCharg1} is described by the $x^5$-dependent analytic superfield $\phi^{+ a}(\zeta, x^5, u)$ which is subjected to the off-shell constraint
\bea
\Big({\cal D}^{++} + H^{++5}\partial_5 +\frac12 \Gamma^{++} \Big) \phi^{+ a} = 0\,,\label{ConstrCchg}
\eea
and possesses the transformation rules
\be
\delta_\lambda \phi^{+ a} = -\frac12\Lambda\,\phi^{+ a}\,, \quad \delta_{\lambda^5} \phi^{+ a} = -\lambda^5 \partial_5\phi^{+ a}\,.
\ee
Its HSS action is written as \cite{Book}
\be
S_{\phi^+} = -\frac12 \int du\,d\zeta^{(-4)} H^{++5}\phi^{+ a}\partial_5\phi^{+}_a\,.  \label{CChact}
\ee
As distinct from the $q^+$ superfield with an infinite number of off-shell component fields, $\phi^{+ a}$ amounts to the finite off-shell set of $8 + 8$ components. It is easy to check that the constraint \p{ConstrCchg} and the action \p{CChact} are superconformally invariant. Their  $\lambda^5$ invariance can be checked using
the constraint \p{ConstrCchg} itself and the transformation law $\delta_{\lambda^5 }H^{++5} = {\cal D}^{++}\lambda^5$. Also, it is straightforward to show that $\partial_5\big[H^{++5}\phi^{+ a}\partial_5\phi^{+}_a\big]$
is a total derivative, so there is no need to integrate over $x^5$ in \p{CChact}. Indeed, using \p{ConstrCchg} and the identity $\partial_5\phi^a \partial_5\phi_a = 0$, one obtains
$$
\partial_{5}\big[H^{++5}\phi^{+ a}\partial_5\phi^{+}_a\big] \simeq \Big[\Big({\cal D}^{++} +\frac12 \Gamma^{++} \Big) \phi^{+ a}\Big] \partial_5\phi_a\,,
$$
where $\simeq$ means ``up to a total derivative''. Applying once more the constraint \p{ConstrCchg} and making use of the identity $\partial_5\phi^a \partial_5\phi_a = 0$, we are convinced that this expression is vanishing.


\subsection{Principal version of Einstein ${\cal N}=2$ supergravity and general matter couplings}
An important problem relevant to any kind of supergravity is how to construct its most general couplings to matter which would extend those known in the rigid supersymmetry. As
was mentioned earlier, the most general ${\cal N}=2$ matter in the rigid case is described by sigma models with general HK bosonic target spaces \cite{AGF}. As was
shown in \cite{BW}, the bosonic target manifolds of ${\cal N}=2$ sigma models in the Einstein supergravity
background are {\it quaternion-K\"ahler} (QK) \cite{BW}, in contrast to the HK ones in the flat
${\cal N}=2$ case \footnote{QK manifolds are $4n$-dimensional Riemannian manifolds with the holonomy group in $Sp(n)\times Sp(1)$. In the QK sigma models in the ${\cal N}=2$ SG background
the $Sp(1)$ curvature $R$ can be normalized
so that  $R =-8n(n+2)\kappa^2$. }.
The latter are described by the analytic superspace action \p{333} involving $2n$
unconstrained hypermultiplet superfields $q^{+a}(\zeta), \;a = 1,\ldots, 2n$, the bosonic components of which parametrize $4n$-dimensional HK manifolds, with an arbitrary interaction
Lagrangian $L^{+4}(q^+, u)$ called the HK potential. Any HK manifold corresponds to a definite HK potential,  and vice versa, any $L^{+4}$ after passing to component fields generates the target bosonic
metric which is guaranteed to be HK \cite{HKpaper,TNpaper}. So the ${\cal N}=2$ hypermultiplet actions provides a powerful method of explicit calculation of diverse HK metric,
and as such it was used, e.g.,  in \cite{TNpaper,GIOT,GiVa}.

The problem of generalization of these hypermultiplet couplings to the case of ${\cal N}=2$ SG  required, first of all, finding out the appropriate density ensuring the invariance
of the analytic HSS integration measure. Indeed, the relevant actions were expected to generalize the rigid ${\cal N}=2$ supersymmetric ones which are defined just as integrals over
the analytic HSS. Unfortunately, there is no way to construct such densities in the framework of the standard ${\cal N}=2$ SG versions based on the compensators with finite
sets of off-shell fields. For instance, using the constrained hypermultiplet with central charge $\phi^a(\zeta, x^5, u))$ one can construct the analytic density $(u^{+a} \phi_a)^2$
with he transformation properties required, $\delta_\lambda\,(u^{+a} \phi_a)^2 = -\Lambda (u^{+a} \phi_a)^2$, but it depends on $x^5$ and so cannot
be used for constructing invariant $4D$ actions. In the case of improved tensor multiplet as a compensator it is also impossible to construct the compensating density
from the basic objects of the relevant action, $(\ell^{++}(\zeta, u), c^{(ik)})$.

An exception is the so called ``principal'' version of ${\cal N}=2$ SG which is based on the choice, as a compensator,
of the unconstrained $q^+$ hypermultiplet with {\it infinitely} many auxiliary fields off shell. It is a completely novel possibility suggested by the HSS approach. It could
not be discovered in the approaches using the component fields or constrained ordinary ${\cal N}=2$ superfields.

This version can be derived in the following way, starting from the version with nonlinear multiplet $N^{++}$ as a compensator. Insert the constraint \p{NonlMflat} into the action with the proper superfield Lagrange multiplier
\bea
S_{\omega, N}  =  \frac12 \int du\, d\zeta^{(-4)} \omega^2 \big[ H^{+4} - {\cal D}^{++}N^{++} - (N^{++})^2\big]. \lb{omegaN}
\eea
To secure superconformal invariance, one is led to ascribe to $\omega$ the transformation property
\bea
\delta \omega = -\frac12 \Lambda \omega\,. \lb{TranOmegaC}
\eea
Now we observe that the action \p{omegaN} is none other than a change of variables in the covariantized $q^{+}$ action
\bea
&& S_{\omega, N}  =  -\frac12 \int du\, d\zeta^{(-4)} q^{+}_i {\cal D}^{++} q^{+ i}\,, \quad q^+_i = (u^+_i - N^{++} u^-_i) \omega\,, \nn \\
&& \omega = u^-_i q^{+i}\,, \;\; N^{++} = \frac{u^+_i q^{+i}}{u^-_j q^{+j}}\,, \quad  \delta q^+_i = -\frac12 \Lambda q^+_i\,. \lb{qactC}
\eea
So the total action of this version of Einstein ${\cal N}=2$ SG can be equivalently written as
\bea
&& S_{N=2}^{prin} = -\frac{1}{\kappa^2} \int du d^4x_A d^4\theta^+_A d^4\theta^-_A\; E^{-1}\,H^{++ 5}H^{-- 5} \nn \\
&& - \frac12 \int du d\zeta^{(-4)}\big[ q^{+}_i \widehat{{\cal D}}^{++} q^{+ i}  - i \frac{\xi}{\kappa}\, (\tau_3)^{ik}{q}^+_i H^{++5} q^+_k\big], \lb{Action}
\eea
where we denoted $\widehat{{\cal D}}^{++}$ the part of ${\cal D}^{++}$ independent of $H^{++ 5}$ and added the term generating in components the
cosmological constant $\sim \xi$.  This term appears after identifying, {\it a l\'a} Scherk and Schwarz, the derivative $\partial_5$ with  the $U(1)$ generator of $SU(2)$
acting on the doublet indices and commuting with supersymmetry, $\partial_5 q^+_j = i\frac{\xi}{\kappa}\, (\tau_3)_j^{\;k} \lambda^5\, q^+_k$. Note that the Maxwell and hypermultiplet parts of the action have wrong signs,
which matches with the role of the relevant multiplets as the compensating ones.

The harmonic superspace approach clearly exhibits the property that the bosonic target space of  sigma models coupled to ${\cal N}=2$ SG is QK \cite{aG16,aG210}. Moreover,
it offers an efficient tool of the explicit calculation of
QK metrics \cite{aG210,IVa}. It was shown in \cite{IVa} that the most general Lagrangian of hypermultiplets in the background of conformal ${\cal N}=2$ SG, one of these hypermultiplet just being a compensator,
provides the QK superfield potential which encodes any bosonic QK metric. It passes into the general HK potential $L^{+4}(q^+, u)$ when the Newton-Einstein coupling constant $\kappa$ trends to zero.
So it is just the principal version of ${\cal N}=2$
Einstein supergravity which admits the most general matter couplings and so
gives rise to the generic QK sigma models in its bosonic sector,
in a complete agreement with the theorem of Bagger and Witten \cite{BW}.

The crucial attractive feature of the principal version is the existence of the analytic density $\omega = u^-_i q^{+i}$ compensating the gauge transformation
of the analytic superspace integration measure.  No such density can be defined in other versions, associated with the matter compensating superfields  possessing finite numbers
of auxiliary fields and subjected to some constraints. Just due to this remarkable property, one can couple to conformal ${\cal N}=2$ SG an arbitrary  ${\cal N}=2$ sigma model.
One introduces $2n$ hypermultiplets
$Q^{+ A}, A = 1, ..., 2n$ having zero weight with respect to ${\cal N}=2$ superconformal group and constructs the following locally ${\cal N}=2$ superconformal action directly
generalizing the rigidly supersymmetric action \p{333}
\bea
&&S_{q, Q} = \frac12\int du\, d\zeta^{(-4)}\big\{ - q^+_i {{\cal D}}^{++} q^{+ i} + (u^-_i q^{+i})^2\big[Q^{+}_A{{\cal D}}^{++} Q^{+ A} \nonumber \\
&& \quad \quad\quad\quad\quad\quad + {\cal L}^{+ 4}(Q^+, v^+, u^-)\big]\big\}, \lb{ActionQK} \\
&& \;v^{+ i} := \frac{q^{+ i}}{u^-_i q^{+i}}\,\nonumber
\eea
(for simplicity, we chose $\xi=0$). The target bosonic geometry associated with this action is QK as distinct from the HK geometry associated with
the action \p{333}. The QK potential ${\cal L}^{+ 4}$, like the HK one, ${L}^{+ 4}$, is an arbitrary function of its arguments and it encodes
the whole set of QK metrics including those without isometries. On the other hand, while coupling conformal SG to the compensators
with finite sets of off-shell components,
the matter-SG couplings prove to be inevitably  restricted just because such compensators are subject to the proper constraints.
For instance, the matter hypermultiplets self-couplings in this case  must reveal some isometries.
More details on the HSS description of QK sigma models and specific examples can be found in \cite{IVa}.
In particular, the general algorithmic scheme of deriving the target QK metrics from the action \p{ActionQK}
was presented there. Also, another convenient trick essentially simplifying the derivation of QK metrics with isometries,
the superfield quotient construction, was worked out.

Note that the central charge $\partial_5$, the inclusion
of which is necessary for gaining a non-zero cosmological constant $\xi$, nontrivially acts on the compensator $q^{+ i}$ in \p{ActionQK} (see \p{Action}), and the invariance of
\p{ActionQK} under $\partial_5$ can be achieved only for some special choices of the QK potential ${\cal L}^{+ 4}$. In other words, in the presence of cosmological constant the
class of admissible target QK metrics gets essentially restricted.

It is also worth noting that in the case of pure Einstein ${\cal N}=2$ SG (without matter couplings) various off-shell SG versions become equivalent via some kind of duality transformations. For instance,
the action of the improved tensor multiplet \p{Impr} can be extended by adding the relevant harmonic constraint to the action with the analytic Lagrange multiplier $\omega'(\zeta)$,
\be
S_{impr} \;\Rightarrow \; S_{impr} + \int du\, d\zeta^{(-4)} \,\omega'\Big({\cal D}^{++} + \Gamma^{++} \Big) L^{++}\,, \quad \delta_\lambda \omega' = 0\,. \lb{imprMod}
\ee
Using more sophisticated change of variables compared to \p{qactC}, the superfields $(\ell^{++}, \omega')$ can also be combined into the hypermultiplet $q^{+ i}{}'$
so that the modified action \p{imprMod} will transform into the covariantized action $\sim q^{+}_i{}'{\cal D}^{++}q^{+ i}{}'$, thus demonstrating that the corresponding ${\cal N}=2$ SG
is also duality-equivalent to the principal version. No such an equivalency holds after switching on couplings to matter. This is because, after integrating out the relevant Lagrange multipliers like $\omega$ or $\omega'$,
the arising constraints on the superfields like $l^{++}$ or $N^{++}$ will be heavily modified by the matter superfields. For instance, making the change of variables \p{qactC} in the action \p{ActionQK},
we transform it to the expression
\bea
&& S^{gen}_{\omega, N} = \frac12 \int du\, d\zeta^{(-4)} \omega^2 \big[ H^{+4} - {\cal D}^{++}N^{++} - (N^{++})^2  + Q^{+}_A{{\cal D}}^{++} Q^{+ A} \nonumber \\
&&  \quad \quad\quad\quad\quad\quad +\, {\cal L}^{+ 4}(Q^+, u^+ - N^{++}u^-, u^-)\big]. \lb{omegaNmod}
\eea
It is obvious that the superfields $Q^{+ A}$ give an essential non-vanishing contribution to the modified constraint on $N^{++}$ obtained by varying with respect to $\omega$ in \p{omegaNmod}.
In the gauge $N^{++} = 0$ this constraint
yields a complicated expression for the harmonic vielbein $H^{+4}$ in terms of matter hypermultiplets.

\setcounter{equation}{0}
\section{Summary and some further problems}
In this short overview of the HSS approach to conformal and Einstein ${\cal N}=2$ supergravities we basically
focused on their symmetry and geometric structures, leaving aside the issues of comparing with the component formulations
and/or those in the standard and projective superspaces (see, e.g., \cite{Howe,N2SgProj}). The main novel advantageous feature of the HSS formulations
(compared to other ones) is that all basic off-shell ${\cal N}=2$ SG multiplets (Weyl multiplet and diverse compensating and matter multiplets) are accommodated
by unconstrained analytic harmonic ${\cal N}=2$ superfields, like it takes place in rigid ${\cal N}=2$ supersymmetry for SYM and hypermultiplet matter multiplets.
This analogy becomes especially striking for the principal
version of Einstein ${\cal N}=2$ SG, which amounts to coupling of ${\cal N}=2$ Weyl multiplet described by unconstrained analytic harmonic vielbeins covariantizing the
harmonic derivative $D^{++}$, to the hypermultiplet compensator described by an unconstrained analytic superfield $q^{+a}$. The local superconformal ${\cal N}=2$ group has the fundamental
realization as the harmonic analyticity-preserving  diffeomorphisms of HSS. The principal version of Einstein
${\cal N}=2$ SG involves an infinite number of auxiliary fields and its invention  is the main outcome of the HSS approach in application to the supergravity
theory. It could not be discovered in any other formulation and ensures the most general ${\cal N}=2$ SG - matter coupling. From mathematical point of view, the
corresponding superfield Lagrangian makes manifest the one-to-one correspondence between local ${\cal N}=2$ supersymmetry and quaternion-K\"ahler sigma models
\cite{BW}, since the basic object of this Lagrangian, ${\cal L}^{+4} (Q^{+}, v^+, u^-)$, is just the unconstrained basic object of the QK geometry, the QK potential.
It is the direct analog of K\"ahler potential $K(\Phi_A, \bar\Phi_B)$ in ${\cal N}=1$ supersymmetry (see eq. \p{Phi}) and HK potential in rigid  ${\cal N}=2$ supersymmetry $L^{+4}(q^+, u^\pm)$ (see eq. \p{333}).
The general Lagrangian of the principal version of ${\cal N}=2$ SG in interaction with matter hypermultiplet superfields given by eq. \p{ActionQK} can serve as an efficient tool
of explicit calculation of the QK metrics, both the already known and yet unknown ones.

The HSS approach bears a close relationship to the
famous twistor theory. Common for both is
an extension of space-time (in twistor theory) and superspace (in
the harmonic superspace approach) by a two-dimensional
sphere $S^2$. In such an extended space the self-dual Yang-Mills
or Einstein equations admit an interpretation as Cauchy-Riemann
conditions associated with some harmonic analyticities, in a close suggestive
analogy with ${\cal N}=2$ matter, SYM and SG theories
in HSS as described above. These striking affinities between
the implications of the harmonic methods in the extended supersymmetries
and in the purely bosonic problems were e.g. used in \cite{HKpaper,aG210} to construct
unconstrained geometric formulations of the hyper-K\"ahler and quaternion-K\"ahler
geometries. These formulations justified the interpretation of the general analytic $q^+$ interaction
Lagrangian $L^{+4}(q^+, u)$ in \p{333} and its ${\cal N}=2$ supergravity analog ${\cal L}^{+4} (Q^{+}, v^+, u^-)$ in \p{ActionQK} as the
fundamental geometric quantities of both types of the complex Riemannian geometries.

Finally, let us sketch several directions in which the HSS method in applications to SG models
could be further developed.

The formulation of ${\cal N}=2$ SG theories and their couplings to matter through unconstrained analytic
superfields with a nice geometric meaning opens a few perspectives which so far were not still investigated in full.
First of all, it should allow for a self-consistent procedure of quantization, like it has been done for ${\cal N}=2$ SYM theory
(see, e.g., reviews \cite{32,review}). In particular, the manifestly analytic superfield propagators and background superfield
method could be constructed and used to study the structure of quantum corrections to the classical actions and the related geometric
objects. Though  ${\cal N}=2$ SG is certainly non-renormalizable by power-counting, the supersymmetry is capable to improve the UV behavior of ${\cal N}=2$ SG models
and it is just the off-shell supefield quantum SG machinery that would allow to check this in a manifestly supersymmetric and gauge invariant fashion.

The same concerns ${\cal N}=(1,0)$ and ${\cal N}=(1,1)$  SG models in six dimension, the HSS formulation of which was pioneered in \cite{Sok2}\footnote{$6D$ HSS was defined in \cite{6D,6D2}.}. It seems interesting
to hybridize these $6D$ HSS SG models with the quantization methods developed for their $6D$ SYM cousins (see a recent review \cite{6DQuant}).

Another area where the HSS approach could be useful is the construction of higher-order  superfield invariants of ${\cal N}=2$ SG, including, e.g., ${\cal N}=2$ conformal SG
extensions of the standard conformal invariant (square of Weyl tensor \cite{FrTs}). To know the precise structure of such higher-derivative SG invariants is of interest for
exploring the relationships with string theory.

Interesting prospective applications of the HSS approach to self-dual supergravities were started in one of the last papers co-authored by V.I. Ogievetsky, ref. \cite{DO}. The closely related new area
of recent uses of ${\cal N}=2, 4D$ HSS approach is the higher-spin business attracting a lot of attention for last years (see. e.g., \cite{Snow} and refs. therein). Surprisingly, it turned out that
the geometrical HSS formulation of the simplest ${\cal N}=2$ Einstein SG presented in Section 5 admits direct generalizations as the off-shell unconstrained analytic superfield formulations
for ${\cal N}=2$ superextensions of the integer higher spin ${\bf s}\geq 3$ theories. At the free level, ${\cal N}=2$ multiplet with a higher spin ${\bf s}$ is described by a triad of harmonic analytic
superfields \cite{BIZ1}
\bea
h^{++\alpha(s-1)\dot\alpha(s-1)}(\zeta, u),\; h^{++\alpha(s-2)\dot\alpha(s-2)}(\zeta, u),\; h^{++\alpha(s-1)\dot\alpha(s-2)+}(\zeta, u) \;({\rm and\,c.c}), \nonumber
\eea
where $\alpha(s) := (\alpha_1 \ldots \alpha_s), \dot\alpha(s) := (\dot\alpha_1 \ldots \dot\alpha_s)$. For ${\bf s}=2$ the superfield content of ${\cal N}=2$ SG is recognized. In \cite{Cubic} the superfield cubic
vertices of interaction of these higher-spin gauge superfields with the matter hypermultiplets were also constructed. Though these theories (and their actions) are still available only at the linearized
level, there is a hope to extend them to the full nonlinear level, closely following the appropriate steps in the HSS formulation of ${\cal N}=2$ SG theories.  In this respect, it is of high importance to construct higher-spin
generalizations of ${\cal N}=2$ conformal supergravity in HSS formulation (at least, at the linearization level).

As a challenge to the HSS approach, there remains the problem of constructing off-shell ${\cal N}=3$ supergravity in terms of the appropriate unconstrained
harmonic potentials. While the nice off-shell formulation of ${\cal N}=3, 4D$ SYM theory (equivalent to ${\cal N}=4$ SYM on shell) in ${\cal N}=3$ HSS
with infinite numbers of auxiliary fields is known for many years \cite{aG7,aG8}, it is still mystery whether some analogous formulation of ${\cal N}=3, 4D$ SG can be invented. The main question is as to what is
 the appropriate off-shell HSS carrier of ${\cal N}=3$ superconformal Weyl multiplet. Once this is understood, the ${\cal N}=3$ SG could be obtained from the superconformal one
via some compensating procedure making use of the proper number of off-shell ${\cal N}=3$ Maxwell multiplets in the background of ${\cal N}=3$ Weyl multiplet\footnote{For on-shell component formulations
of ${\cal N}=3$ SG and vector ${\cal N}=3$ multiplets see, e.g., a recent preprint \cite{N3On} and refs. therein.}. It is not excepted that solving this
ambitious problem will require some new geometrical ideas beyond those incorporated within the standard ${\cal N}=3$ HSS approach (see, e.g., discussion in \cite{GIO}).

\section*{Acknowledgements}
It is pleasure for me to thank S. James Gates Jr. and the Editors of the {\it Handbook of Quantum Gravity} for the kind invitation to present this contribution. I am indebted to Emery Sokatchev for long-lasting
fruitful collaboration on the HSS approach and related topics. I thank Konstantinos Koutrolikos for a technical assistance. This work was supported in part
by the Ministry of Education of the Russian Federation, project FEWF-2020-003.

\end{document}